\documentclass[12pt]{elsarticle}

\usepackage{graphicx}%
\usepackage{multirow}%
\usepackage{amsmath,amssymb,amsfonts}%
\usepackage{amsthm}%
\usepackage{mathrsfs}%
\usepackage{appendix}%
\usepackage{xcolor}%
\usepackage{textcomp}%
\usepackage{manyfoot}%
\usepackage{booktabs}%
\usepackage{algorithm}%
\usepackage{algorithmicx}%
\usepackage{algpseudocode}%
\usepackage{listings}%
\usepackage{subcaption}%
\usepackage{tabularx}
\usepackage{adjustbox}
\usepackage{lineno}

\newcommand{\bb}[1]{
\textbf{#1}
}

\newcommand{\bo}[1]{
\boldsymbol{#1}
}

\newcommand{\R}[1]{
\textcolor{red}{#1}
}

\begin{document}

\begin{frontmatter}



\title{Held-to-Maturity Accounting: Hidden Losses and Systemic Fragility\tnoteref{tnote1}} 

\tnotetext[tnote1]{Given one of the co-authors is working for J.P. Morgan Chase, we hence are required to put the following disclaimer: The information, views, and opinions set forth in this publication are those of the authors and are in no ways sponsored, endorsed, or related to the business of J.P. Morgan Chase \& Co. (‘J.P. Morgan’). J.P. Morgan does not warrant the publication’s completeness or accuracy and makes no representations regarding the use of the information set forth in this publication. In no event shall J.P. Morgan be liable for any direct, indirect, special, punitive, or consequential damages, including loss of principal and/or lost profits, even if notified of the possibility of such damages. Nothing in this publication is intended to be an advertisement or offer for any J.P. Morgan service.}
 

\author[addr1]{Dongxu Li} 
\ead{dli36@stevens.edu}
\author[addr1]{Steve Y. Yang\corref{cor1}} 
\ead{syang14@stevens.edu}
\cortext[cor1]{Corresponding author.}
\affiliation[addr1]{organization={School of Business, Stevens Institute of Technology}, addressline={1 Castle Point on Hudson}, city={Hoboken}, postcode={NJ 07030}, country={USA}}

\begin{abstract}
This paper studies how accounting-induced opacity affects systemic risk in the banking system. Motivated by the buildup of unrealized losses during recent monetary tightening episodes, we develop a model in which banks hold correlated interest rate sensitive assets and are connected through an endogenous interbank network. We show that Held-To-Maturity (HTM) accounting amplifies systemic risk by delaying loss recognition, allowing losses to accumulate and become synchronized across institutions. When losses are realized under liquidity stress, the interaction of correlated exposures and network linkages generates nonlinear increases in defaults. These findings suggest that accounting regimes are not neutral from a financial stability perspective, as they influence both the buildup and propagation of systemic risk, particularly in environments with common asset exposures and strong interconnectedness.

\end{abstract}


\date{March 21, 2026} 
\end{frontmatter}



\section{Introduction}
\label{sec:introduction}

Financial stability in the banking system is increasingly tested during periods of monetary policy tightening, when rising interest rates erode asset values and place pressure on bank balance sheets. While this channel has been widely studied, recent episodes highlight the importance of accounting practices in shaping how these risks materialize. The failure of Silicon Valley Bank provides a salient example, where large unrealized losses on securities classified as Held-To-Maturity remained unrecognized until liquidity pressures forced their realization. More generally, accounting regimes that permit the deferral of unrealized losses can obscure the true economic value of bank assets, creating a gap between reported and market-based measures of financial health. Such opacity allows losses to accumulate during normal periods and emerge abruptly under liquidity stress, thereby amplifying uncertainty and increasing the potential for systemic instability. Understanding how monetary tightening interacts with accounting-induced opacity is therefore central to assessing the resilience of the banking system.

A central institutional feature underlying these concerns is the rapid expansion of Held-To-Maturity (HTM) securities on bank balance sheets in the post-pandemic period. Following large-scale liquidity injections and a prolonged low interest rate environment, banks significantly increased their holdings of long-duration securities, particularly U.S. government bonds and mortgage-backed securities. As interest rates rose sharply during the subsequent monetary tightening cycle, the market value of these securities declined substantially. However, under HTM accounting, these unrealized losses are not reflected in reported earnings or regulatory capital, provided that banks assert the intent and ability to hold these assets to maturity. Recent evidence suggests that banks with greater exposure to interest rate risk and weaker capital positions were more likely to reclassify securities into HTM, indicating a strategic use of accounting discretion to manage reported balance sheet conditions~\citep{granja2023bank}. At the same time, aggregate HTM holdings have increased markedly across the banking sector, while exposures to similar interest rate–sensitive assets have contributed to rising comovement in bank asset values \citep{jiang2024monetary}.\footnote{Using U.S. bank-level data, we document a substantial increase in HTM securities holdings in recent years (see Table~\ref{Securities Ratio Stats}), alongside elevated asset return correlations across large banks (see Table~\ref{returnCorrelation}) and a positive association between HTM exposure and aggregate correlation measures (see Table~\ref{regressionResults}). }

Taken together, these patterns give rise to a fundamental puzzle for financial stability. While HTM accounting may smooth reported earnings and reduce short-term balance sheet volatility, it may also facilitate the buildup of correlated, unrecognized losses across banks. When liquidity conditions tighten—through deposit outflows or funding stress—these latent losses may be realized in a synchronized manner, potentially amplifying system-wide stress. This raises an important question: can accounting-induced opacity, while stabilizing in normal times, paradoxically increase systemic fragility during periods of monetary tightening?

To address this puzzle, the paper proposes a mechanism through which accounting-induced opacity interacts with asset correlation and financial interconnectedness to amplify systemic risk. Under HTM accounting, valuation losses on long-duration assets are not immediately recognized, allowing negative shocks to accumulate on bank balance sheets over time. When banks hold similar interest rate–sensitive securities, these latent losses become highly correlated across institutions, consistent with evidence on common asset exposures and return comovement in the banking sector ~\citep{duarte2021fire,jiang2024monetary}.
 Crucially, these accumulated losses are not realized continuously but may instead be triggered simultaneously when banks face liquidity pressures, such as deposit outflows or funding shocks. This synchronization transforms gradual valuation declines into discrete, system-wide balance sheet shocks, resembling fire-sale dynamics and liquidity-driven amplification mechanisms documented in prior studies~\citep{cifuentes2005liquidity,caballero2013fire}.

The systemic implications of this mechanism are further amplified by the structure of interbank networks. A large body of literature shows that financial interconnectedness can propagate and magnify shocks through contagion effects~\citep{eisenberg2001systemic,glasserman2016contagion,houy2020defaulting}.
 In this setting, the combination of correlated asset exposures and synchronized loss realization increases the likelihood that multiple institutions experience distress simultaneously, weakening their ability to meet interbank obligations and amplifying default cascades. This interaction between hidden losses, correlation, and network contagion implies that the timing of loss recognition is not neutral: accounting rules that delay recognition may reduce observed volatility in normal times but can endogenously generate larger and more concentrated shocks during stress periods. As a result, the mechanism provides a unified explanation for how accounting practices, asset commonality, and network structure jointly contribute to nonlinear systemic instability in a tightening monetary environment.

To formalize this mechanism, the paper develops a dynamic agent model of the interbank system in which heterogeneous banks interact through lending, borrowing, and portfolio allocation decisions. The framework builds on the literature on financial networks and endogenous formation of interbank connections, where links arise from decentralized interactions rather than being imposed exogenously~\citep{lux_emergence_2015,liu_interbank_2020,jackson_systemic_2020}. In the model, each bank manages a stylized balance sheet consisting of external assets, interbank exposures, and deposit liabilities, and makes borrowing and lending decisions subject to leverage constraints. Through repeated interactions and adaptive behavior, banks form interbank credit relationships endogenously, generating a network structure that resembles the empirically observed core-periphery topology of financial systems~\citep{soramaki2007topology,jackson_systemic_2020}.

Within this environment, banks are exposed to stochastic and potentially correlated asset returns, reflecting common exposures to interest rate–sensitive securities. The model incorporates a clearing mechanism in the spirit of~\cite{eisenberg2001systemic} and its extension ~\cite{RogersVeraart2013}, allowing for limited liability and partial recovery in the event of default. More broadly, the framework is consistent with recent approaches that use agent-based and network-based models to study systemic risk and financial stability~\citep{tasca2016market,bookstaber2018agent}. This structure enables us to simulate the joint evolution of bank balance sheets, interbank connections, and default outcomes under different shock scenarios. By introducing correlated asset shocks and varying the timing of loss realization, the framework provides a tractable setting to evaluate how accounting regimes influence the buildup and propagation of systemic risk.

The analysis yields three main findings that highlight the financial stability implications of accounting-induced opacity. First, the model shows that higher correlation in bank asset returns can lead to a nonlinear increase in systemic risk. As asset return correlations rise, reflecting common exposures to interest rate–sensitive securities, the likelihood of simultaneous distress across institutions increases sharply, resulting in a disproportionate rise in bank defaults. Quantitatively, increasing asset return correlation from 0.8 to 0.9 leads to a nearly fivefold increase in the average number of bank defaults, rising from approximately 12 to over 58 institutions in the simulated system.\footnote{See Table~\ref{app:asset_correlation} for default statistics under varying correlation assumptions. } This result is consistent with the literature emphasizing the role of common exposures and fire-sale spillovers in amplifying systemic risk~\citep{cifuentes2005liquidity,duarte2021fire}.

Second, the timing of loss recognition plays a critical role in shaping systemic outcomes. Holding the total magnitude of asset losses fixed, the model shows that when losses are realized in a single period—as in the case of HTM accounting under liquidity stress—the banking system experiences a sharp increase in defaults. In contrast, when the same losses are distributed over time, as under a mark-to-market (MTM) regime, the system is significantly more resilient. In particular, breaking a single large shock into two or three smaller shocks reduces the average number of defaults from 12.03 to 1.68 and 0.08, respectively.\footnote{See Table~\ref{app:asset_correlation} for the comparison between single-shock and multiple-shock scenarios. } This finding underscores that accounting rules affect not only the measurement of losses but also their temporal distribution, which in turn influences banks’ ability to adjust leverage and manage risk dynamically.

Third, the interaction between correlated asset exposures and interbank network structure further amplifies systemic fragility. In an interconnected system, distress at one institution can propagate through credit linkages, leading to cascade effects that disproportionately affect highly connected banks. When shocks are both correlated and synchronized, these network effects become particularly pronounced, consistent with theoretical and empirical evidence on contagion in financial networks~\citep{eisenberg2001systemic,glasserman2016contagion,jackson_systemic_2020}.

These findings suggest that accounting-induced opacity can alter not only the level of risk in the banking system but also its distribution and timing, thereby playing a central role in the emergence of systemic instability during periods of monetary tightening. Recent evidence by \cite{begenau2026q} shows that banks accumulated substantial unrealized losses during the prolonged low-interest-rate period and demonstrates that these losses contributed to the fragility observed during the 2023 banking turmoil. Their analysis primarily focuses on individual bank balance sheets and the interaction between interest-rate risk, depositor runs, and bank solvency. In contrast, our paper investigates whether the HTM accounting framework itself may generate systemic externalities that extend beyond individual institutions. Specifically, we study how delayed loss recognition can propagate through interbank networks, amplify correlated asset exposures, and increase systemic risk during periods of monetary tightening. While \cite{begenau2026q} document the existence and magnitude of unrealized losses, our analysis examines the systemic consequences of hiding those losses within the banking system. The results suggest that accounting-induced opacity may affect not only the resilience of individual banks but also the transmission of shocks across financial institutions, thereby linking accounting choices to financial stability outcomes.

Overall, this paper contributes to the existing literature on financial stability by identifying a novel mechanism through which accounting practices can amplify systemic risk in the banking system. In particular, the paper shows that the use of Held-To-Maturity (HTM) accounting transforms gradual valuation losses into latent balance sheet distortions that accumulate over time and are only realized under liquidity stress. When banks hold similar long-duration assets, such as government securities, this delayed loss recognition leads to highly synchronized and correlated shocks across institutions. From a financial stability perspective, this mechanism implies that risks may remain hidden during normal periods but materialize abruptly during stress events, thereby increasing the likelihood of joint defaults and systemic disruptions. This finding highlights the important role of accounting design in shaping the timing and magnitude of systemic risk.

Secondly, the paper contributes to the literature by developing a dynamic, agent model of the interbank system that explicitly links balance sheet dynamics, endogenous network formation, and correlated asset exposures. Unlike existing models that typically rely on exogenous or static network structures, the framework allows banks to form and adjust credit relationships through a learning process, generating a core-periphery network consistent with empirical observations. This modeling approach is particularly relevant for financial stability analysis, as it captures how interconnectedness and endogenous behavior jointly determine the propagation of shocks. By embedding correlated asset returns within this network structure, the model provides a tractable and empirically grounded setting to examine how localized balance sheet vulnerabilities can escalate into system-wide instability.

Finally, the paper contributes to the policy debate on accounting regimes and financial stability by demonstrating that the timing of loss recognition plays a critical role in shaping systemic outcomes. The results show that, even when the total magnitude of asset losses is unchanged, a mark-to-market (MTM) approach, where losses are recognized more gradually, significantly reduces the incidence of bank defaults compared to HTM accounting. From a financial stability standpoint, earlier and more continuous loss recognition allows banks to adjust leverage, rebalance portfolios, and manage interbank exposures before risks become systemic. This suggests that accounting transparency can serve not only as an informational tool but also as a macroprudential mechanism that mitigates contagion risk by desynchronizing shocks across institutions.

The remainder of the paper is organized as follows. Section~\ref{sec:background} provides institutional background on Held-To-Maturity accounting and presents key stylized facts on the evolution of bank balance sheets, HTM securities holdings, and asset return comovement. Section~\ref{sec:mechanism} introduces the economic mechanism linking accounting-induced opacity, correlated asset exposures, and systemic risk. Section~\ref{sec:model} presents the model framework, describing the balance sheet structure of banks, the endogenous formation of interbank networks, and the key features of the simulation design. Section~\ref{sec:validation} discusses model calibration and validation, demonstrating that the simulated system reproduces key empirical features of interbank networks and bank balance sheet dynamics. Section~\ref{sec:results} reports the main results, quantifying the effects of asset correlation and the timing of loss recognition on systemic risk, and evaluating the implications of HTM versus mark-to-market accounting. Section~\ref{sec:implications} discusses policy implications for accounting design and financial stability. Section~\ref{sec:conclusion} concludes.

\section{Empirical Evidence}
\label{sec:background}
\subsection{HTM Accounting and Monetary Tightening}
\label{sub:htm_accounting}
Held-To-Maturity (HTM) accounting allows banks to record certain fixed-income securities at amortized cost rather than fair market value, provided that they assert both the intent and the ability to hold these assets until maturity. This accounting treatment reduces short-term earnings volatility by shielding balance sheets from fluctuations in market prices. However, it also creates a divergence between the reported book value and the underlying economic value of assets, particularly when interest rates change significantly. As a result, unrealized losses on long-duration securities may remain unrecognized in financial statements for extended periods.

This feature becomes especially relevant in a tightening monetary environment. Rapid increases in interest rates reduce the market value of fixed-income securities, such as U.S. Treasury bonds and mortgage-backed securities, which constitute a large share of bank portfolios. While these valuation losses are immediately reflected under mark-to-market accounting, they remain largely hidden under HTM classification unless the assets are sold. Consequently, banks with substantial HTM holdings may accumulate significant unrealized losses without corresponding adjustments to reported capital or earnings~\citep{laux2010did,huizinga2012bank}.

Recent empirical evidence suggests that banks actively adjust their accounting classifications in response to these incentives. In particular, banks with greater exposure to interest rate risk, weaker capitalization, or higher reliance on uninsured deposits are more likely to reclassify securities into HTM during periods of rising rates~\citep{granja2023bank}. At the same time, monetary tightening can trigger deposit outflows as investors seek higher yields in alternative assets, such as money market funds, thereby increasing liquidity pressures on banks. When such pressures force asset sales, previously unrealized losses in HTM portfolios become realized, potentially leading to sharp declines in bank capital and confidence. Recent studies document that the gap between the market and book value of bank assets has widened significantly during the recent tightening cycle, highlighting the extent of these hidden losses~\citep{jiang2024monetary}.

Overall, HTM accounting can provide stability in normal times by smoothing reported earnings, but it may also obscure the buildup of interest rate risk on bank balance sheets. In a tightening environment, this opacity can delay the recognition of losses and contribute to sudden adjustments under stress, thereby interacting with liquidity conditions to amplify financial fragility.

\subsection{Stylized Facts}
\label{sub:stylized_facts}
\begin{figure}[ht!]
    \centering
    \includegraphics[width=0.8\textwidth]{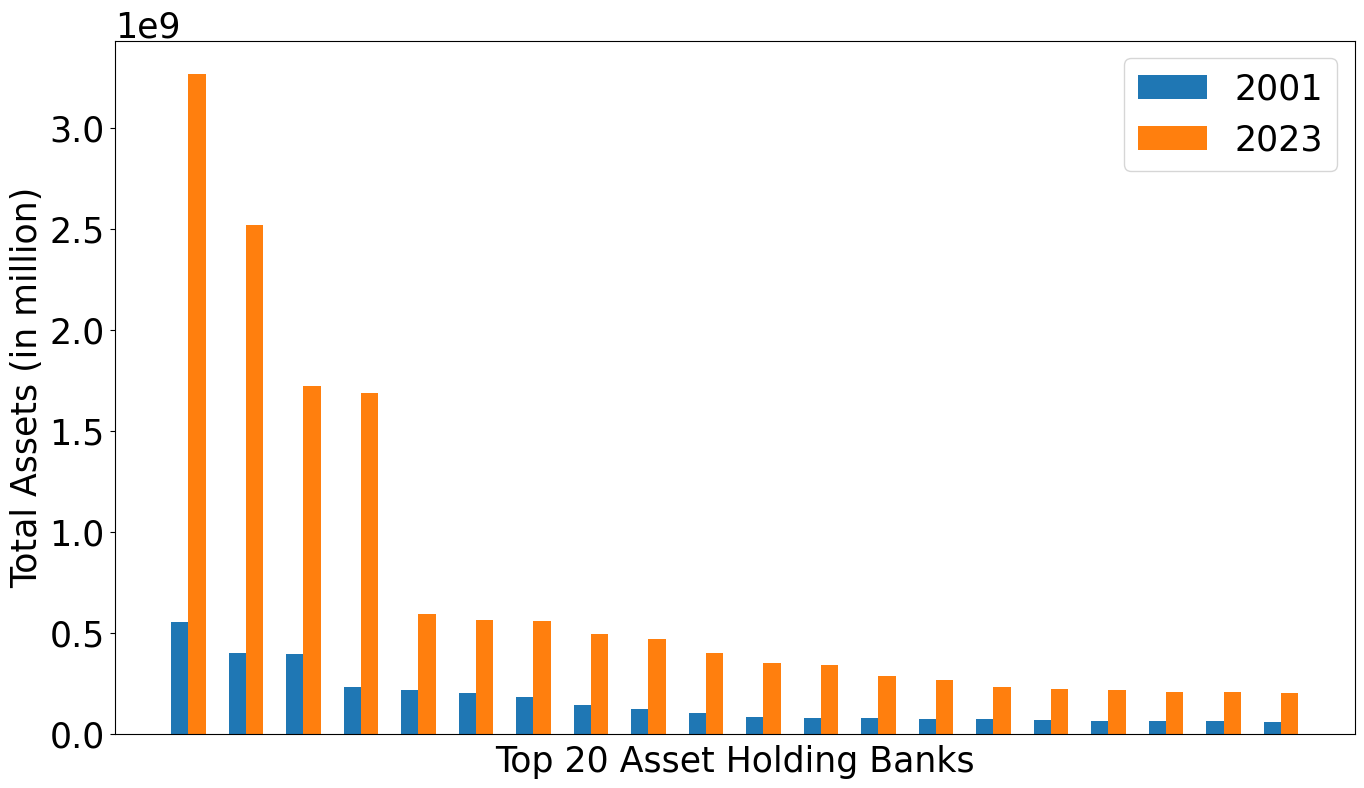}
    \caption{Top Asset Holding banks}
    \label{topAssetHolding}
\end{figure}

We begin by documenting structural changes in the U.S. banking system over the past two decades. Table \ref{numberOfBanks} shows a steady decline in the number of banks alongside a significant increase in asset concentration among the largest institutions. Assets have grown significantly from 2001 to 2023; the largest bank holds about 3.5 trillion of assets in the year 2023 (see Figure \ref{topAssetHolding}), and the top 4 largest banks accumulate more assets that dwarf those of smaller banks. We witness the consolidation of banks' assets, where the five largest banks in the U.S. controlled about 23\% total assets in 2001, 36\%  in 2007, and 44\%  in 2015. This consolidation implies that a smaller set of banks now accounts for a larger share of total assets, increasing the importance of interconnectedness and concentration for systemic risk. Such developments are consistent with the literature emphasizing the role of network centralization and the systemic importance of large financial institutions~\citep{soramaki2007topology,jackson_systemic_2020}.

\begin{table}[ht!]
\centering
\begin{tabular}{ l| c c c c c c c c} 
\hline 
Year & 2002 & 2005 & 2008 & 2011 & 2014 & 2017 & 2020 & 2023 \\
\hline
\# Of Banks & 8689 & 8294 & 7805 & 6985 & 6794 & 5911 & 5171 & 4681 \\
\hline
\end{tabular}
\caption{Number of banks covered by FFIEC}
\label{numberOfBanks}
\end{table}

\begin{figure}[ht]
     \centering
     \begin{subfigure}[b]{0.45\textwidth}
         \centering
         \includegraphics[width=\textwidth]{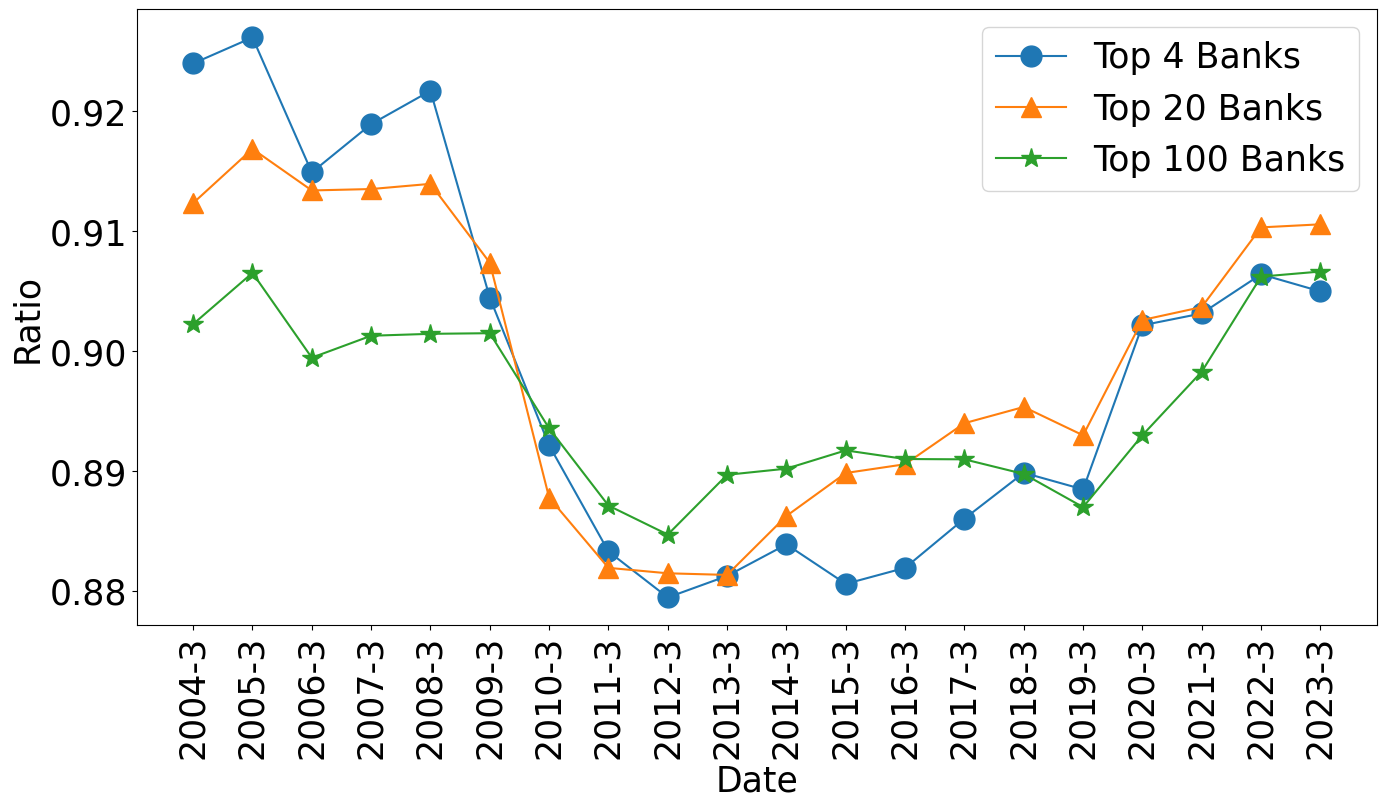}
         \caption{Total Liabilities / Total Assets}
         \label{fig: liability}
     \end{subfigure}
     \hfill
     \begin{subfigure}[b]{0.45\textwidth}
         \centering
         \includegraphics[width=\textwidth]{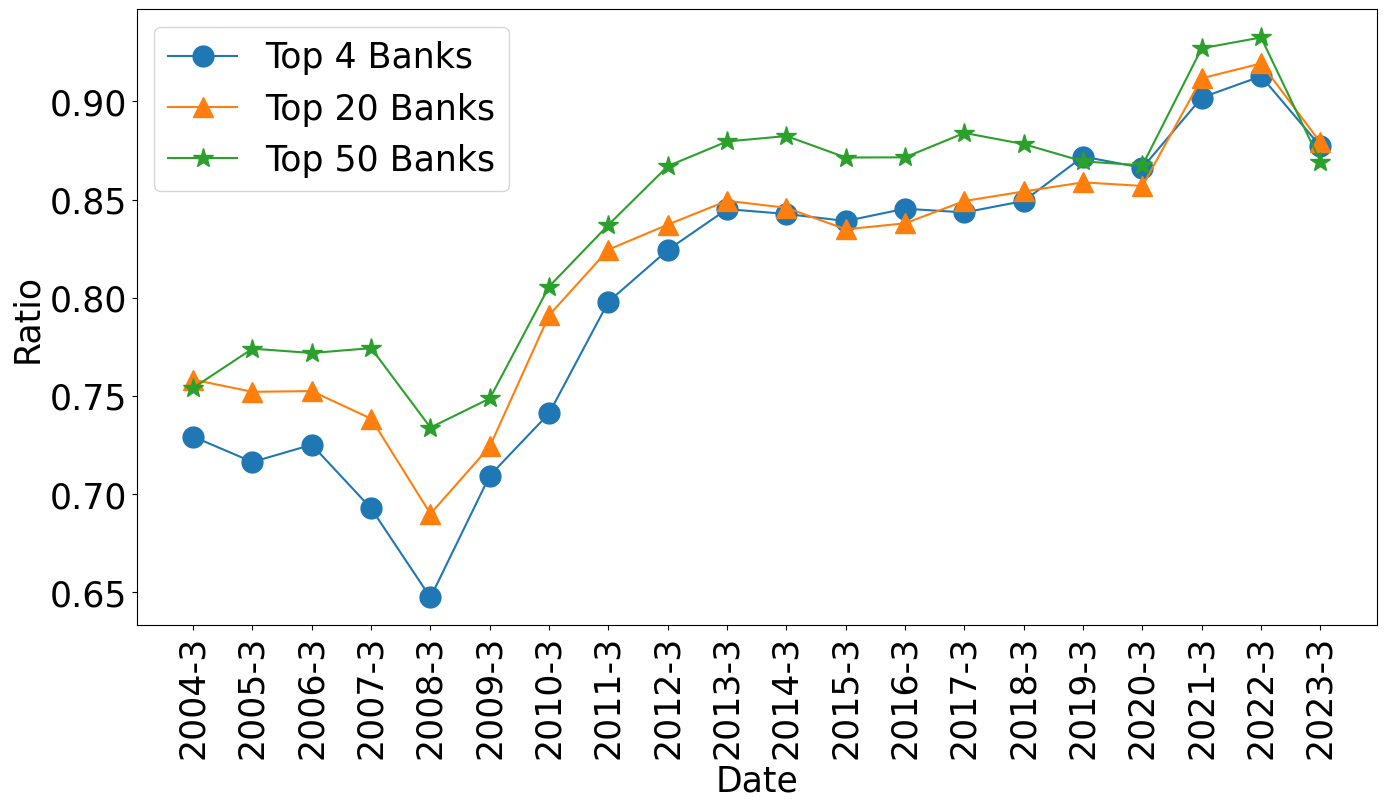}
         \caption{Total Deposit / Total Liabilities}
         \label{fig: deposit}
     \end{subfigure}
     \hfill
     \begin{subfigure}[b]{0.45\textwidth}
         \centering
         \includegraphics[width=\textwidth]{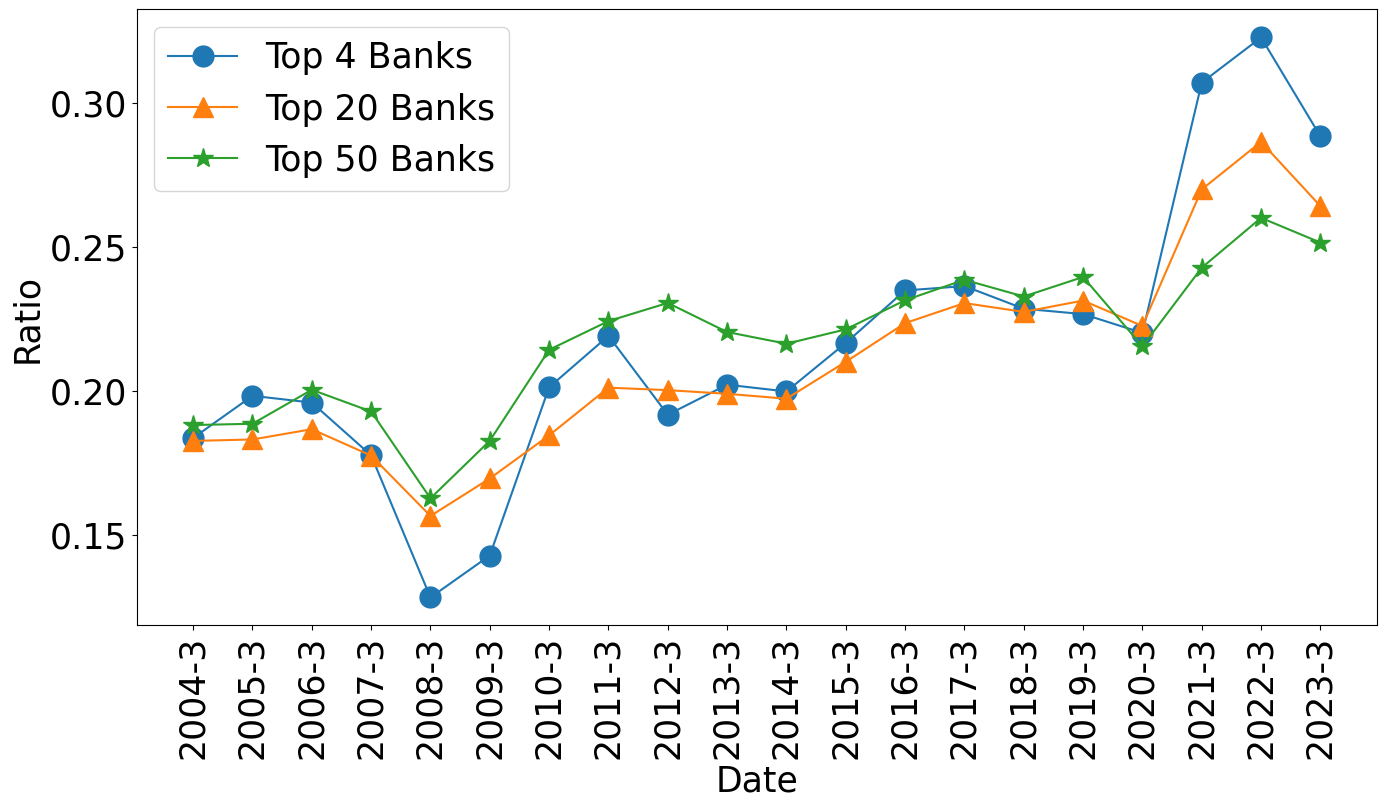}
         \caption{Total Securities / Total Assets}
         \label{fig: securities}
     \end{subfigure}
     \hfill
     \begin{subfigure}[b]{0.45\textwidth}
         \centering
         \includegraphics[width=\textwidth]{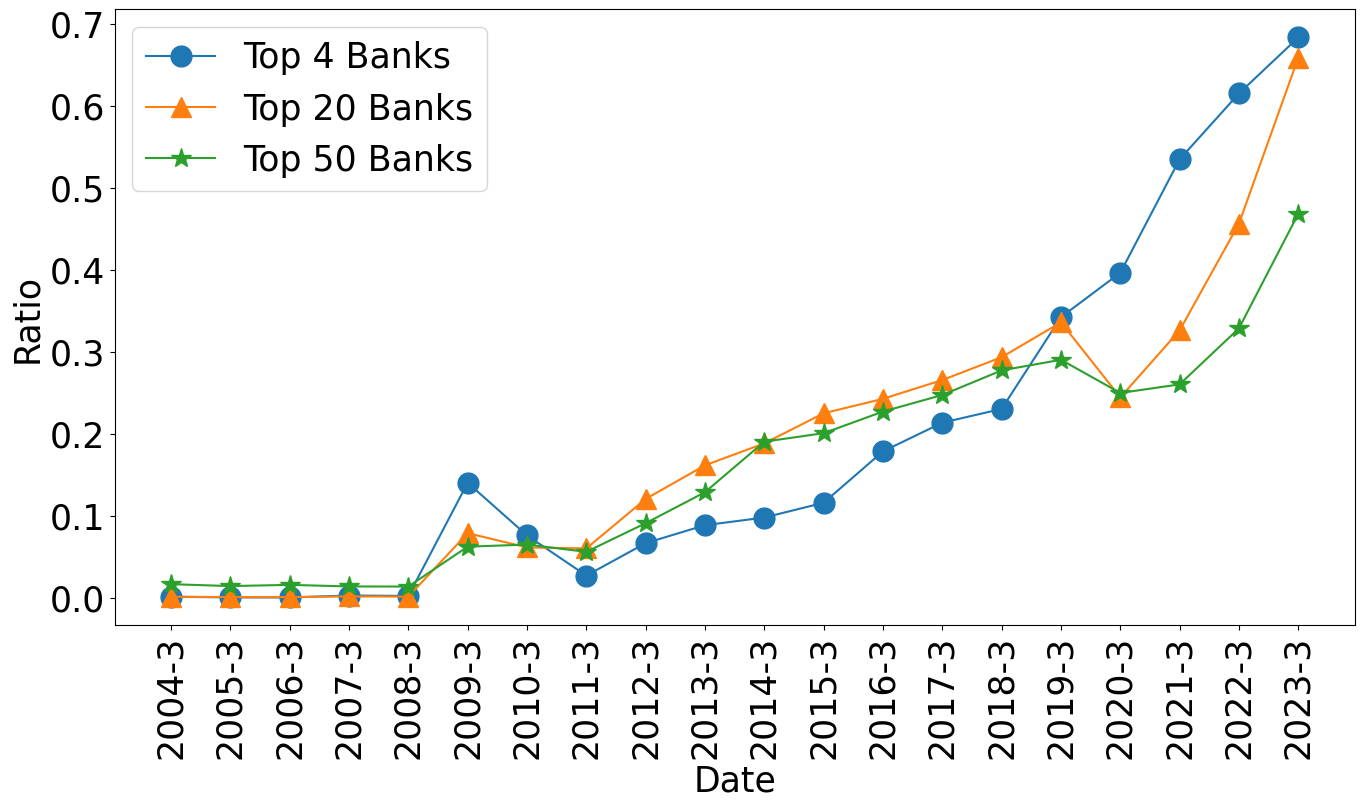}
         \caption{Held to Maturity / Total Securities}
         \label{fig: HTMsecurities}
     \end{subfigure}
        \caption{Trends in Banks' Balance Sheet}
        \label{fig:balancesheetStats}
\end{figure}

Second, we observe pronounced shifts in bank balance sheet composition in the post-pandemic period. As illustrated in Figure \ref{fig:balancesheetStats}, banks experienced a substantial increase in deposit funding following periods of monetary expansion, accompanied by a notable reallocation of assets toward securities holdings. In particular, deposits became the dominant funding source for banks, while lending activity remained relatively subdued, leading banks to invest excess liquidity in fixed-income securities. At the same time, leverage ratios remained relatively stable, indicating that these adjustments occurred primarily through changes in asset composition rather than capital structure. This combination of high deposit funding and increased exposure to interest rate–sensitive securities suggests that banks became more vulnerable to valuation losses during subsequent monetary tightening episodes.\footnote{Figure~\ref{fig:balancesheetStats} reports average balance sheet ratios for the top 4, 20, and 50 asset-holding banks. Deposits rose sharply following liquidity injections, with the deposit-to-liabilities ratio exceeding 90\% in March 2022 before declining to approximately 85\% in March 2023, consistent with deposit outflows into higher-yield alternatives such as money market funds. Over the same period, the share of securities in bank portfolios increased by roughly 10\% for large banks and 5\% for smaller banks, reflecting weak loan demand and portfolio reallocation toward fixed-income assets. A substantial portion of these securities was classified as HTM, with HTM ratios rising to around 70\% for large banks and 45\% for smaller banks. Moreover, more than 80\% of these holdings consist of U.S. government securities, implying significant exposure to interest rate risk.} These patterns align with existing evidence on how banks respond to monetary policy and liquidity conditions~\citep{duffie2018financial,demirgucc2021banking}.

Third, a key feature of this reallocation is the substantial increase in Held-To-Maturity (HTM) securities. Figure~\ref{fig:balancesheetStats} and Table~\ref{Securities Ratio Stats} show that, while the overall share of securities in bank portfolios remains relatively stable, the fraction classified as HTM has risen markedly for the most significant 50 banks. From an accounting perspective, this shift is consequential. Under U.S. GAAP, HTM securities are measured at amortized cost, and unrealized gains and losses are not reflected in earnings or regulatory capital, in contrast to Available-for-Sale (AFS) securities, which are marked to market through other comprehensive income (OCI). As interest rates increase, this accounting treatment creates a divergence between the book value and market value of bank assets, allowing unrealized losses to accumulate without immediate impact on reported capital~\footnote{While SILICON VALLEY BANK (SVB)
189 had a securities ratio of 0.572, the highest among the regional banks, and both
190 FIRST REPUBLIC BANK and SILICON VALLEY BANK had the highest
191 HTM Securities Ratios of alarmingly 0.886 and 0.792 in March 2022, which are
192 way above the average ratio (Appendix Table \ref{HTMratio Rank}).}. Prior studies have documented that such accounting discretion can influence bank behavior and obscure underlying financial conditions~\citep{laux2010did,huizinga2012bank}.

\begin{table}[ht]
\centering
\resizebox{1.0 \columnwidth}{!}{%

    \begin{tabular}{lrr}
    \hline
    {\bb{Securities Ratio}} &  March 2022 &  March 2023 \\
    \hline
    mean &                  0.245 &                  0.221 \\
    std  &                  0.143 &                  0.138 \\
    min  &                  0.029 &                  0.013 \\
    25\%  &                  0.164 &                  0.139 \\
    50\%  &                  0.226 &                  0.203 \\
    75\%  &                  0.290 &                  0.254 \\
    max  &                  0.758 &                  0.773 \\
    \hline
    \end{tabular}

    \begin{tabular}{lrr}
    \hline
    {\bb{HTM Securities Ratio}} &  March 2022 &  March 2023 \\
    \hline
    mean &                                     0.238 &                                     0.352 \\
    std  &                                     0.241 &                                     0.283 \\
    min  &                                     0.000 &                                     0.000 \\
    25\%  &                                     0.018 &                                     0.038 \\
    50\%  &                                     0.147 &                                     0.347 \\
    75\%  &                                     0.392 &                                     0.568 \\
    max  &                                     0.886 &                                     0.999 \\
    \hline
    \end{tabular}
}%

\caption{Securities Ratio / HTM Securities Ratio Summary Statistics}
\label{Securities Ratio Stats}
\end{table}

Fourth, we document a notable increase in asset return comovement across banks. Table \ref{returnCorrelation} reports pairwise correlations of daily stock returns for large U.S. banks, showing that average correlations rise from approximately 0.76 to 0.89 in the post-COVID period. Given that bank equity returns reflect underlying asset performance, this increase suggests that banks have become more exposed to common risk factors, particularly interest rate movements. To further quantify system-wide comovement, we construct an asset-weighted correlation measure. The regression results in Appendix \ref{app:asset_correlation} indicate that both the securities ratio and the HTM ratio are positively associated with this aggregate correlation, implying that greater exposure to fixed-income securities, and especially HTM-classified assets, is linked to higher correlation in asset returns across banks. This evidence is consistent with the literature on common asset holdings and systemic risk, which highlights how overlapping portfolios can amplify the transmission of shocks~\citep{greenwood2015vulnerable,duarte2021fire}.

    \begin{table}[ht]
    \centering
\begin{adjustbox}{width=\textwidth}
\begin{tabular}{ccccccccccc}
\toprule
{pre-COVID\bb{post-COVID}} &   BAC &               BK &                C &               GS &              JPM &               MS &              PNC &              TFC &              USB &              WFC \\
\midrule
BAC &       &  0.63(\bb{0.87}) &  0.83(\bb{0.93}) &  0.86(\bb{0.89}) &  0.85(\bb{0.95}) &  0.82(\bb{0.89}) &  0.83(\bb{0.92}) &   0.8(\bb{0.91}) &  0.85(\bb{0.93}) &  0.73(\bb{0.91}) \\
BK  &  0.24 &                  &  0.59(\bb{0.84}) &  0.52(\bb{0.82}) &  0.54(\bb{0.86}) &  0.55(\bb{0.84}) &  0.62(\bb{0.81}) &    0.6(\bb{0.8}) &  0.63(\bb{0.83}) &   0.54(\bb{0.8}) \\
C   &   0.1 &             0.25 &                  &  0.78(\bb{0.89}) &  0.87(\bb{0.92}) &  0.86(\bb{0.87}) &   0.78(\bb{0.9}) &  0.75(\bb{0.89}) &   0.78(\bb{0.9}) &  0.73(\bb{0.88}) \\
GS  &  0.03 &              0.3 &             0.11 &                  &  0.79(\bb{0.89}) &  0.83(\bb{0.92}) &  0.71(\bb{0.84}) &  0.72(\bb{0.81}) &  0.75(\bb{0.82}) &  0.67(\bb{0.84}) \\
JPM &   0.1 &             0.32 &             0.05 &              0.1 &                  &  0.85(\bb{0.89}) &  0.85(\bb{0.93}) &   0.78(\bb{0.9}) &  0.81(\bb{0.92}) &  0.68(\bb{0.89}) \\
MS  &  0.07 &             0.29 &             0.01 &             0.09 &             0.04 &                  &  0.76(\bb{0.84}) &  0.72(\bb{0.82}) &  0.75(\bb{0.84}) &  0.69(\bb{0.86}) \\
PNC &  0.09 &             0.19 &             0.12 &             0.13 &             0.08 &             0.08 &                  &  0.82(\bb{0.93}) &  0.85(\bb{0.92}) &  0.71(\bb{0.88}) \\
TFC &  0.11 &              0.2 &             0.14 &             0.09 &             0.12 &              0.1 &             0.11 &                  &  0.82(\bb{0.93}) &  0.68(\bb{0.89}) \\
USB &  0.08 &              0.2 &             0.12 &             0.07 &             0.11 &             0.09 &             0.07 &             0.11 &                  &   0.74(\bb{0.9}) \\
WFC &  0.18 &             0.26 &             0.15 &             0.17 &             0.21 &             0.17 &             0.17 &             0.21 &             0.16 &                  \\
\bottomrule
\end{tabular}
\end{adjustbox}

\caption{Daily Stock Return Correlation of 10 large U.S. banks \\ The upper triangular matrix presents the correlations before the pandemic on the left and the correlations after the pandemic in the bracket, and the lower triangular matrix shows the increase of the correlation after the pandemic.}
    \label{returnCorrelation}
    \end{table}

These empirical patterns point to an environment in which banks are increasingly exposed to similar sources of risk while simultaneously reporting balance sheets that may not fully reflect underlying market valuations. In such a setting, shocks to asset values—particularly those driven by interest rate changes—are more likely to be both correlated and delayed in recognition. This combination suggests that losses may materialize in a synchronized manner under stress, providing a natural link to the mechanism explored in the next section.

\section{Related Literature}
\label{sec:mechanism}

This section outlines related literature that supports the economic mechanism through which accounting-induced opacity, common asset exposures, and financial interconnectedness jointly contribute to systemic fragility. The mechanism operates through three related channels. First, HTM accounting delays the recognition of valuation losses and allows economic losses to accumulate on bank balance sheets. Second, because many banks hold similar interest rate–sensitive securities, these hidden losses are likely to be highly correlated across institutions. Third, once losses are realized under stress, the interbank network can propagate and amplify those shocks through funding withdrawals, balance-sheet contractions, and counterparty distress. This framework connects the paper to a broad literature in finance and accounting on bank opacity, delayed loss recognition, liquidity creation, and systemic risk.

\subsection{Hidden Loss Accumulation}

The first strand of literature supports the mechanism with respect to delayed loss recognition. Under HTM accounting, unrealized losses on fixed-income securities are not immediately reflected in reported earnings or regulatory capital so long as banks maintain the intent and ability to hold those securities to maturity. This treatment can smooth reported performance in normal times, but it also widens the gap between book values and market values when interest rates rise sharply. The accounting literature emphasizes that financial reporting choices affect how bank risk is perceived and priced, and that discretion in recognition and classification can alter both market discipline and managerial incentives \citep{BeattyLiao2014, laux2010did, huizinga2012bank}.

More recent work highlights that delayed loss recognition is not merely a measurement issue but can interact with capital constraints to increase fragility. In particular, \cite{BegenauBigioMajerovitzVieyra2024} show that delayed loss accounting can amplify banking system vulnerability when book values diverge from economic values. In a tightening cycle, long-duration securities decline in value as interest rates rise, yet these losses remain largely hidden under HTM classification unless assets are sold. This delay matters because reported balance-sheet strength may overstate resilience precisely when funding conditions deteriorate. Empirical evidence shows that unrealized losses became economically significant during the recent tightening episode and contributed to banking system fragility \citep{jiang2024monetary}. Related work suggests that accounting classifications and limited hedging may be used to manage reported balance sheet risk without eliminating underlying exposure \citep{GranjaEtAl2024}. 

\subsection{Correlation Amplification}

The second strand of literature speaks about the mechanism of correlation amplification through common asset holdings. When many banks invest in similar long-duration securities, shocks to interest rates generate valuation losses that are strongly synchronized across institutions. This idea connects to the literature on common asset exposures and systemic risk. \cite{greenwood2015vulnerable} show that overlapping portfolios increase vulnerability to shocks, while \cite{ duarte2021fire} demonstrate how fire-sale spillovers transmit distress across institutions through shared asset positions.

HTM accounting introduces an important temporal dimension to this mechanism. Losses are not recognized continuously as market prices evolve but instead accumulate across institutions and may be realized simultaneously under stress. This feature transforms gradual valuation declines into discrete system-wide shocks. The mechanism is also consistent with broader theories of opacity and liquidity creation. \cite{Jungherr2018} shows that bank opacity can weaken market discipline and increase risk-taking, while \cite{AcharyaThakor2016} highlight that liquidity creation and leverage can contribute to systemic fragility when underlying risks are not transparently measured. In this setting, HTM accounting and common asset exposures jointly increase both the magnitude and synchronization of losses across banks.

\subsection{Network Contagion}

The third strand of literature talks about the mechanism that operates through the interbank network. Banks are connected through lending and borrowing relationships, so that distress at one institution can propagate to others. The classic framework of \cite{eisenberg2001systemic} shows how interbank obligations can generate default cascades, while \cite{RogersVeraart2013} incorporate recovery frictions that further amplify contagion.

More recent work emphasizes that network structure determines how shocks propagate through the system. \cite{DenbeeJulliardLiYuan2021} show that key institutions play a central role in liquidity provision, while \cite{CabralesGottardiVegaRedondo2017} highlight the trade-off between risk sharing and contagion in financial networks. These studies suggest that interconnectedness can stabilize the system under small shocks but amplify systemic risk when shocks are large or correlated. This insight is consistent with \cite{glasserman2016contagion}, who show that contagion becomes particularly severe when shocks are correlated across institutions.

In the present setting, the combination of delayed loss recognition and correlated exposures implies that multiple banks may become impaired simultaneously. When this occurs, the network's capacity to absorb shocks is weakened, and contagion effects are amplified. Recent evidence from the Silicon Valley Bank episode supports this mechanism, showing that unrealized losses and funding fragility contributed to cross-bank spillovers \citep{ChoiGoldsmithPinkhamYorulmazer2023}. The interbank network therefore acts as an amplification layer that transforms hidden and synchronized losses into system-wide instability.

\section{Model Framework}
\label{sec:model}

This section presents a stylized model of the banking system designed to capture the interaction between accounting-induced opacity, correlated asset exposures, and interbank contagion. The framework integrates endogenous network formation, balance sheet dynamics, and a network-based default mechanism, while remaining parsimonious to isolate the key economic forces.

\subsection{Environment}

We consider a banking system consisting of $n$ banks indexed by $\mathcal{N} = \{1,2,...,n\}$. Banks are represented as nodes in a directed network, where edges correspond to interbank liabilities. Let $\bo{L} \in \mathbb{R}^{n \times n}$ denote the liability matrix, where $L_{ij}$ represents the nominal liability of bank $i$ to bank $j$. We assume $L_{ij} \geq 0$ and $L_{ii}=0$.

Each bank holds a portfolio of external assets and interbank claims. Specifically, bank $i$'s total assets consist of its portfolio $E_i$ and incoming interbank payments $\sum_j L_{ji}$. The portfolio $E_i$ captures all assets outside the interbank system, including securities, loans, and other holdings. We decompose the portfolio into an investable component $I_i$ and a non-investable component $\Bar{I}_i$, where $\Bar{I}_i$ represents capital reserves required by regulation. This distinction allows us to capture the idea that only part of the balance sheet is actively exposed to market risk, while the remainder serves as a buffer against losses.

\begin{equation}
\begin{split}
    \text{Total assets} &= E_i + \sum_j L_{ji} \\ 
    \text{Portfolio } E_i &= I_i + \Bar{I}_i
\end{split}
\end{equation}

On the liability side, bank $i$ finances its activities through interbank borrowing and deposits. Total liabilities are therefore given by $\sum_j L_{ij} + D_i$, and equity $E_i$ is defined as the residual claim:

\begin{equation}
\begin{split}
    \text{Total liability} = \sum_j L_{ij} + D_i \\
    \text{Total assets} - \text{Total liability} = E_i
\end{split}
\end{equation}

Deposits constitute the primary funding source, consistent with the empirical evidence in Section 2. As a result, banks operate with substantial leverage even in the absence of interbank borrowing.

Interbank transactions affect balance sheets in economically intuitive ways. Borrowing increases both assets and liabilities without affecting equity, thereby raising leverage. Lending reallocates assets across counterparties but does not change total assets. Debt repayment reduces both assets and liabilities, with corresponding adjustments to equity through interest payments. These dynamics are central to understanding how shocks propagate through the system.

\subsection{Borrowing and Lending Decisions}

Credit relationships are borrower-initiated, consistent with evidence from interbank markets \citep{lux_emergence_2015}. Banks decide both how much to borrow and from whom.

\textit{Borrowing quantity.} Banks borrow subject to a leverage constraint. In each episode, a bank issues borrowing requests until it reaches a maximum leverage ratio or exhausts a fixed number of requests. Unlike models driven by liquidity shocks, borrowing here is motivated by investment opportunities: banks borrow at a relatively low rate and invest in higher-yielding risky assets.

\textit{Counterparty selection.} Each bank $i$ forms beliefs about the likelihood of obtaining funding from other banks. These beliefs are summarized by a probability vector $\bo{P}_i = [p_{i1},...,p_{in}]$, where $p_{ij}$ represents the probability that bank $j$ will extend credit to bank $i$. The borrowing target bank $x$ is then drawn from a categorical distribution:

\begin{equation}
\begin{split}
    \bo{P} = [p_1, p_2,...,p_n] \mid \bo{\alpha} &\sim \text{Dirichlet Distribution}(n, \bo{\alpha}) \\ \\
    X = x \mid \bo{P} &\sim \text{Categorical Distribution}(N, \bo{P})
\end{split}    
\end{equation}

With symmetric initial beliefs ($\alpha_j=1$), banks initially approach all counterparties with equal probability:

\begin{equation}
\begin{split}
    \mathbb{E}[p_j] = \frac{1}{n-1}
\end{split}
\end{equation}

Beliefs evolve through learning. After each episode, banks update their beliefs based on observed loan approvals $\bo{C}=[c_1,...,c_n]$. Using the conjugacy of the Dirichlet prior, the posterior distribution is:

\begin{equation}
\begin{split}
    \bo{P} \mid \bo{C}, \bo{\alpha} &\sim \text{Dirichlet Distribution}(n, \bo{\alpha} + \bo{C}) \\ \\
    \bo{\alpha} &:= \bo{\alpha} + \bo{C} 
\end{split}
\end{equation}

This updating rule implies that banks gradually shift their borrowing toward counterparties that are more likely to approve loans, generating endogenous network structure.

To avoid excessive concentration of beliefs, we introduce a memory decay mechanism that rescales the concentration parameters:

\begin{equation}
\begin{split}
    \bo{\alpha} := \frac{\bo{\alpha}}{\left\| \bo{\alpha} \right\|_1} \cdot (n-1) \cdot Q 
\end{split}
\end{equation}

This ensures that banks continue to explore new counterparties, preventing the network from becoming fully locked-in.

\subsection{Network Formation}

The combination of borrowing decisions and belief updating generates an endogenous interbank network. Banks with higher perceived reliability attract more borrowing requests and become more central over time.

At the same time, banks face similar investment opportunities and regulatory constraints, leading to correlated asset choices \citep{jackson_systemic_2020, elliott_systemic_2021}. These common exposures play a key role in systemic risk, as they make banks vulnerable to the same shocks. In particular, fire-sale dynamics can depress asset values across institutions simultaneously \citep{cifuentes2005liquidity, duarte2021fire}.

In our model, at every episode, the interbank borrowing rate is set to be a constant rate $R_f$; after the borrowing and lending activities, banks invest the rest of their portfolio in risky assets which will grow at stochastic rates $\bo{R} \in \mathbb{R}^{n \times 1}$, with the mean returns $\bo{\mu} \in \mathbb{R}^{n \times 1}$ and standard deviation of the returns $\bo{\sigma} \in \mathbb{R}^{n \times 1}$. We could simulate correlated risk asset returns $\bo{R}$, given the correlation matrix $\bo{\Sigma} \in \mathbb{R}^{n \times n}$ of the returns and a standard Gaussian vector $\bo{Z} \in \mathbb{R}^{n \times 1 }$ where $\bo{Z} \sim \mathcal{N}(\bo{0}, \bo{I})$:

\begin{equation}\label{riskyReturn}
\begin{split}
    \bo{\Sigma}' &= \text{diag}(\bo{\sigma}) \cdot \bo{\Sigma} \cdot \text{diag}(\bo{\sigma}) \\ \\
    \bo{\Sigma}' &= \bo{L} \bo{L}^T \\ \\
    \bo{R} &= \bo{\mu} + \bo{L} \bo{Z}
\end{split}
\end{equation}

This formulation generates correlated returns across banks, capturing exposure to common risk factors such as interest rate movements.

To study systemic crises, we introduce adverse shocks that simultaneously affect all banks. During shock periods, returns become:

\begin{equation}\label{shockModel}
\begin{split}
    \bo{R} &= -\lvert \Delta_{\bo{\sigma}} \cdot \bo{L} \bo{Z} \rvert
\end{split}
\end{equation}

This specification captures sudden drops in asset values and spikes $\Delta_{\bo{\sigma}}$ in volatility, consistent with crisis episodes.

\subsection{Market Clearing and Defaults}

To determine how shocks propagate through the system, we adopt the clearing framework of \cite{eisenberg2001systemic}, extended by \cite{RogersVeraart2013} to incorporate partial recovery.

We consider an interbank system defined by $(\bo{L}, \bo{E}, \kappa_e, \kappa_l)$, where $\bo{L} \in \mathbb{R}^{n \times n}$ is the liability matrix, $\bo{E} = [E_1,...,E_n]$ denotes banks' portfolios of external assets, and $\kappa_e, \kappa_l \in (0,1]$ represent recovery rates on external assets and interbank claims, respectively. The total nominal liabilities of each bank are given by $\Bar{L} = \bo{L}\bo{1}$, where $\Bar{L}_i = \sum_j L_{ij}$ is the total obligation of bank $i$ to all other banks.

Following \cite{eisenberg2001systemic}, the clearing mechanism is based on three key assumptions: (i) limited liability, so that no bank can pay more than its available resources; (ii) priority of debt repayment, meaning that available assets are used to satisfy obligations before equity holders receive any residual value; and (iii) proportional repayment, whereby creditors receive payments in proportion to their claims when a bank defaults.

A clearing vector $L^* \in [0,\Bar{L}]$ specifies the realized payments of all banks in the system, where $L^*_i$ represents the total payment made by bank $i$ to its creditors. The clearing vector is defined as a fixed point of the mapping $\Phi: \mathbb{R}^n \rightarrow \mathbb{R}^n$, i.e., $L^* = \Phi(L^*)$, where $\Phi$ determines each bank's feasible payment given incoming payments from its counterparties.

\begin{equation} \label{clearingVector}
    \begin{split}
            \Phi(L)_i= 
\begin{cases}
    \Bar{L}_i,      & \text{if } \Bar{L}_i \leq E_i + \sum^n_{j=1} L_j \pi_{ji}\\
    \kappa_e E_i + \kappa_l \sum^n_{j=1} L_j \pi_{ji}   & \text{otherwise}
\end{cases}
    \end{split}
\end{equation}

To interpret Equation \ref{clearingVector}, note that $\sum^n_{j=1} L_j \pi_{ji}$ represents the total payment received by bank $i$ from other banks, where $\pi_{ji}$ is the fraction of bank $j$'s total liabilities owed to bank $i$. The total available resources of bank $i$ are therefore the sum of its external assets $E_i$ and incoming interbank payments.

If these resources exceed total liabilities $\Bar{L}_i$, the bank is solvent and repays its obligations in full. Otherwise, the bank defaults and distributes its available resources across creditors after applying recovery rates $\kappa_e$ and $\kappa_l$ to external and interbank assets, respectively.

The relative liability matrix $\bo{\Pi} \in \mathbb{R}^{n \times n}$ is defined as:

\begin{equation}
    \begin{split}
            \pi_{ij}= 
\begin{cases}
   \frac{L_{ij}}{\Bar{L}_i}      & \text{if } \Bar{L}_i > 0\\
    0   & \text{otherwise}
\end{cases}
    \end{split}
\end{equation}

Each element $\pi_{ij}$ represents the share of bank $i$'s total liabilities owed to bank $j$. This matrix determines how payments are distributed across counterparties and plays a central role in the propagation of losses.

The clearing vector $L^*$ is computed using the Greatest Clearing Vector Algorithm (GA) proposed by \cite{RogersVeraart2013}. Once $L^*$ is obtained, we update balance sheets as follows:

\begin{equation} \label{updateBalanceSheet}
    \begin{split}
        L^* &= \text{GA}((\bo{L}, \bo{E}, \kappa_e, \kappa_l)) \\  \\
        \bo{L}' &:= \bo{0} \\ \\
        \bo{E}' &= (\Pi^T L^* + \bo{E} - L^*) \odot \mathbb{I}_{\{L^* \geq \Bar{L}\}}
    \end{split}
\end{equation}

In the above equation \ref{updateBalanceSheet}, the Genetic Algorithm (GA) algorithm is applied to calculate the clearing vector; the debts are all cleared, and the liability matrix becomes a zero matrix. The term $\Pi^T L^*$ captures incoming payments from other banks, while $L^*$ represents outgoing payments. The difference $\Pi^T L^* - L^*$ therefore corresponds to net interbank cash flows. The updated portfolio $\bo{E}'$ reflects these net flows combined with initial assets, conditional on solvency as indicated by the indicator vector $\mathbb{I}_{\{L^* \geq \Bar{L}\}}$.

After clearing, banks may still become insolvent if their equity is negative. This is captured by:

\begin{equation}
    \begin{split}
        \bo{E}' &= \bo{E}' \odot \mathbb{I}_{\{\bo{E}' -\bo{D} \leq 0\}} \\ \\
        \bo{D}' &= \bo{D}' \odot \mathbb{I}_{\{\text{solvent banks}\}}
    \end{split}
\end{equation}

This step ensures that banks with negative equity are identified as defaulting institutions. In such cases, remaining assets are used to repay depositors, while shareholders are fully wiped out.

Overall, the clearing mechanism links balance sheet conditions and network structure. Losses at one bank reduce payments to its creditors, which in turn weakens their balance sheets and may trigger further defaults. This recursive process captures the endogenous propagation of shocks through the interbank network and forms the basis for our analysis of systemic risk.

\section{Calibration and Validation}
\label{sec:validation}

\begin{figure}[ht]
     \centering
     \includegraphics[width=0.8\textwidth]{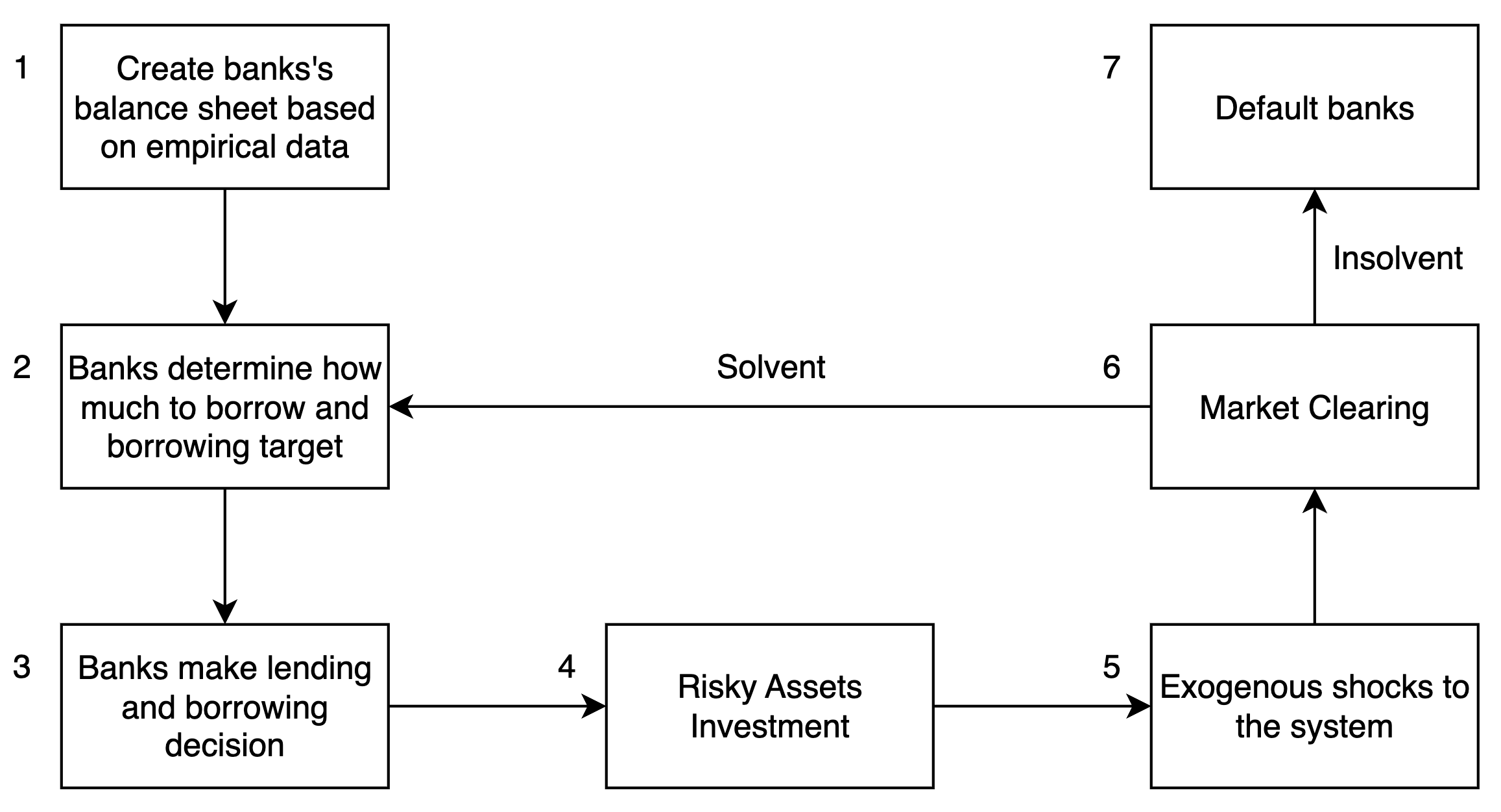}
     \caption{Model Flowchart}
     \label{fig:flowchart}
\end{figure}

This section describes the calibration of model parameters and evaluates the model’s ability to replicate key empirical patterns documented in Section 2. The calibration is guided by observed balance sheet characteristics, asset return dynamics, and interbank connectivity. Key features and summary statistics are documented during the following multi-step process (see Figure~\ref{fig:flowchart}). 

\subsection{Calibration Strategy}

The model is simulated with $N = 100$ banks, corresponding to the largest institutions in the empirical sample. This choice reflects the concentration of assets documented in Figure~\ref{topAssetHolding} and ensures that the model captures the systemic importance of large banks.

Model parameters are summarized in Table~\ref{hyperparameter}. Balance sheet parameters are calibrated to match the empirical distributions reported in Section~\ref{sec:background}. In particular, the share of securities and the proportion of HTM securities are chosen to align with Table~\ref{Securities Ratio Stats}, while the deposit-to-liability ratio reflects the high reliance on deposit funding observed in Figure~\ref{fig:balancesheetStats}.

\begin{table}[ht]
\centering
\begin{tabular}{ l r } 
Hyper-parameters  & Base Model \\
 \hline
Leverage Ratio Upper Bound $\overline{LR}$   & 10.00 \\
Capital Reserve as a ratio of deposit & 0.20 \\
Maximum Number of Borrowing Requests $\overline{NB}$ & 10.00   \\
Size of the borrow as a ratio of Equity      & 1.00  \\
Federal Fund Rate $R_f$                      & 0.04 \\
Portfolio Return Rate $\mu$                  & 0.10 \\
Portfolio Return Volatility $\sigma$         & 0.18 \\
Portfolio Returns Correlation $\rho$         & 0.80 \\
Interbank Asset recovery rate $\kappa_l$     & 0.90 \\
Portfolio recovery rate $\kappa_e$           & 0.50 \\
\hline
\\
\end{tabular}
\caption{Hyperparameter of the base model}
\label{hyperparameter}
\end{table}

The asset return process is calibrated to match the volatility and correlation structure observed in the data. The volatility parameter $\sigma$ captures interest rate–induced valuation changes, while the correlation parameter $\rho$ is chosen to match the increase in asset return comovement documented in Table~\ref{returnCorrelation}.

Network formation is governed by the parameter $\theta$, which controls the concentration of interbank exposures. This parameter is selected to generate a network structure consistent with the concentration patterns observed in the data.

\subsection{Validation of Stylized Facts}

We assess the model’s ability to replicate key empirical features along three dimensions: balance sheet structure, asset distribution, and network topology.

First, the model reproduces key balance sheet characteristics across banks. Table~\ref{assetDistribution} compares the distribution of assets in the simulated model with the empirical distribution. The model closely matches both the level and dispersion of bank asset sizes, capturing the concentration of assets among large institutions.

\begin{table}[ht]
\centering

\begin{tabular}{lrrrrrrr}
\toprule
{} &   Mean &    Std &    Min &    25\% &    50\% &    75\% &     Max \\
\midrule
Step 1:        &  1.937 &  4.875 &  0.192 &  0.289 &  0.475 &  1.531 &  34.957 \\
Step 100:      &  1.846 &  4.646 &  0.187 &  0.280 &  0.450 &  1.461 &  33.558 \\
Step 200:      &  1.833 &  4.556 &  0.191 &  0.294 &  0.460 &  1.496 &  32.201 \\
Step 300:      &  1.868 &  4.655 &  0.183 &  0.293 &  0.463 &  1.516 &  33.678 \\
Step 400:      &  1.792 &  4.476 &  0.176 &  0.278 &  0.444 &  1.467 &  32.019 \\
Step 500:      &  1.799 &  4.519 &  0.159 &  0.273 &  0.446 &  1.437 &  32.583 \\ \\ 
\hline
Empirical Data &  1.925 &  4.846 &  0.192 &  0.287 &  0.469 &  1.519 &  34.767 \\
\bottomrule
\end{tabular}

\caption{Simulated Asset Distribution vs. Empirical Distribution (Assets values in 100 million USD)}
\label{assetDistribution}
\end{table}

\begin{table}[ht]
\centering

\begin{tabular}{lrrrrrrr}
\toprule
{} &    Mean &    Std &    Min &     25\% &     50\% &     75\% &     Max \\
\midrule
Step 1:        &  10.943 &  1.679 &  5.375 &  10.370 &  11.750 &  11.958 &  12.521 \\
Step 100:      &  11.203 &  1.309 &  5.629 &  10.912 &  11.771 &  12.025 &  12.353 \\
Step 200:      &  11.079 &  1.194 &  5.457 &  10.672 &  11.491 &  11.847 &  12.268 \\
Step 300:      &  11.054 &  1.068 &  6.021 &  10.782 &  11.390 &  11.673 &  12.169 \\
Step 400:      &  10.747 &  1.048 &  5.348 &  10.470 &  11.055 &  11.422 &  11.943 \\ 
Step 500:      &  10.576 &  1.051 &  5.065 &  10.268 &  10.844 &  11.251 &  11.661 \\\\
\hline
Empirical Data &  10.573 &  3.034 &  4.603 &   8.763 &  10.215 &  11.585 &  21.514 \\
\bottomrule
\end{tabular}

\caption{Simulated Leverage Distribution vs Empirical Distribution}
\label{leverageDistribution}
\end{table}

Second, the model replicates the leverage distribution observed in the data. Table~\ref{leverageDistribution} reports leverage ratios in both the empirical sample and the simulated model. The model-generated leverage distribution aligns closely with the data, indicating that the calibration successfully captures banks’ funding structure and risk exposure.

\begin{figure}[ht]
     \centering
     \includegraphics[width=0.8\textwidth]{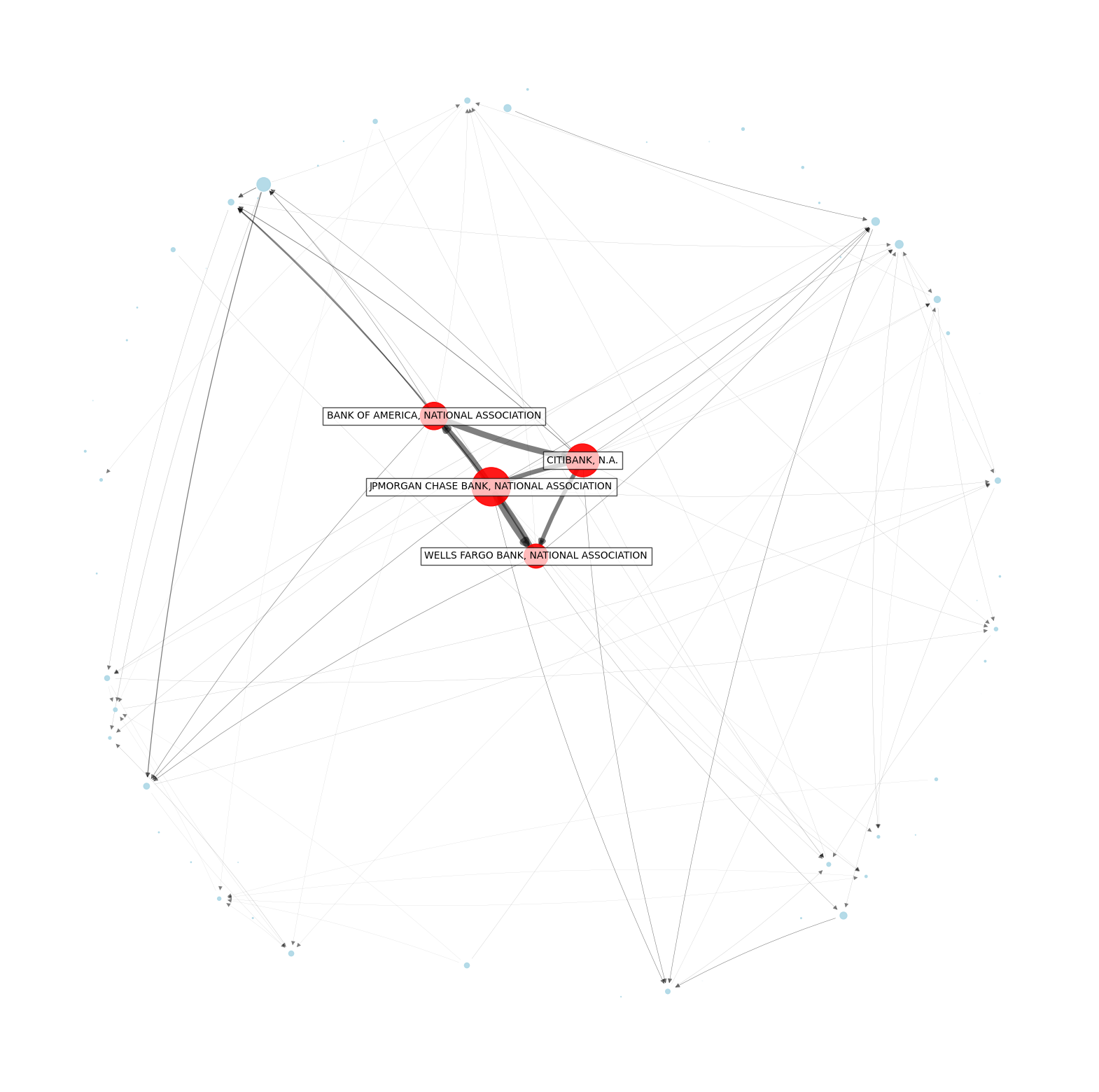}
     \caption{Essential difference in the asset value between HTM and MTM}
     \label{fig: Lnetwork}
\end{figure}

Third, the model generates realistic network structures. Figure \ref{fig: Lnetwork} illustrates the simulated interbank network, which exhibits a core-periphery structure consistent with empirical evidence~\citep{anderson2019bank}. A small number of highly connected banks emerge endogenously and play a central role in interbank activity.

\begin{figure}[ht]
    \centering
    \includegraphics[width=0.8\textwidth]{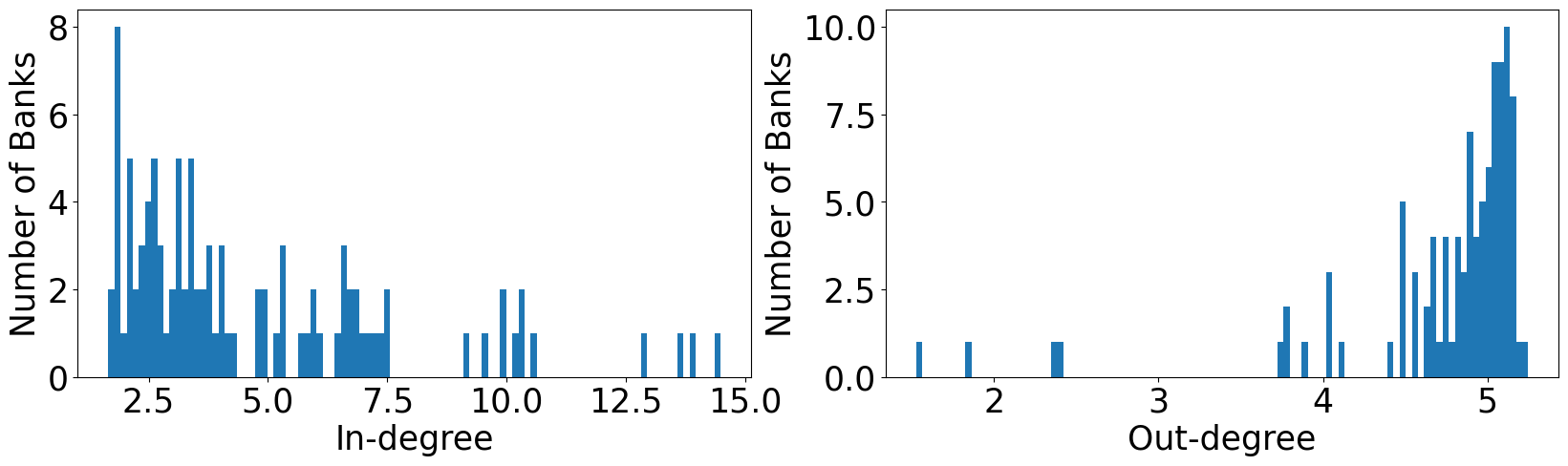}
    \caption{In-degree and Out-degree Histogram}
    \label{inOutDegree}
\end{figure}

\begin{figure}[ht]
    \centering
    \begin{subfigure}[b]{0.495\textwidth}
        \centering    
        \includegraphics[width=\textwidth]{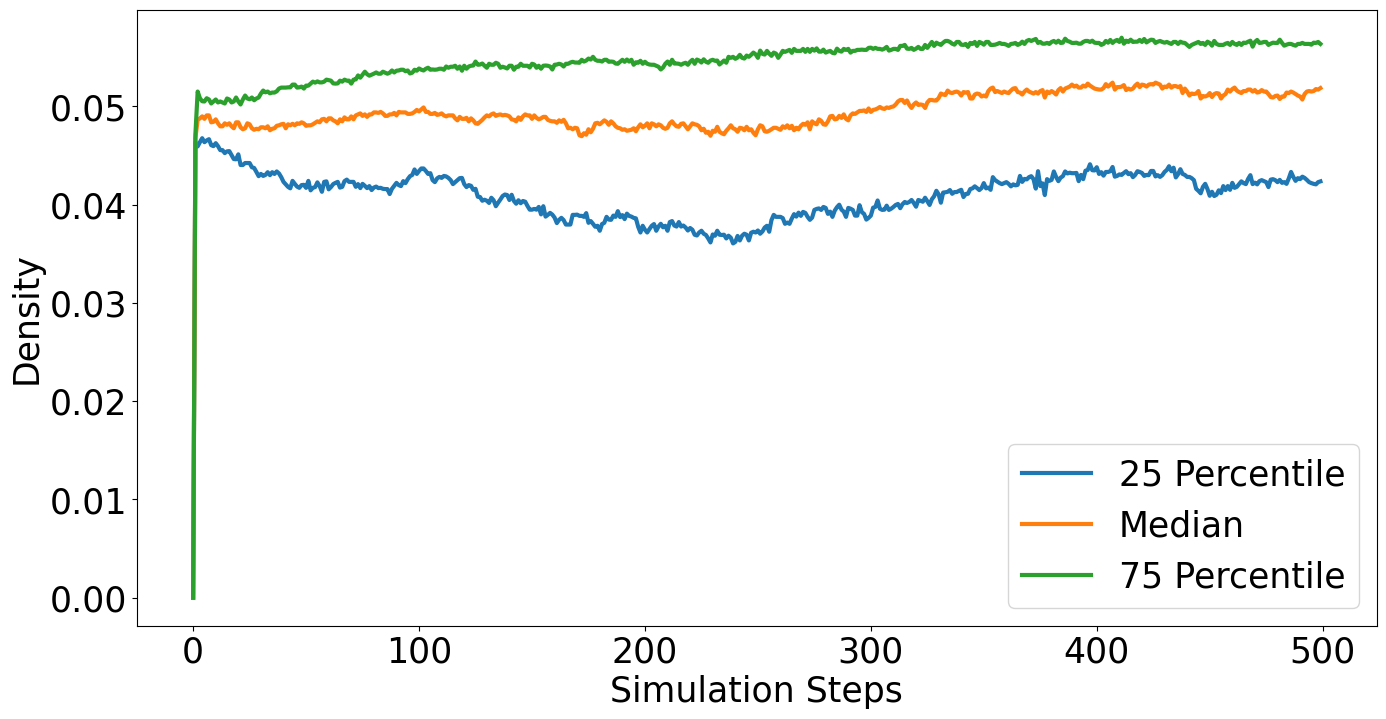}
         \caption{Credit Network Density}
         \label{networkDensity}
    \end{subfigure}
    \hfill
    \begin{subfigure}[b]{0.495\textwidth}        
    \centering
        \includegraphics[width=\textwidth]{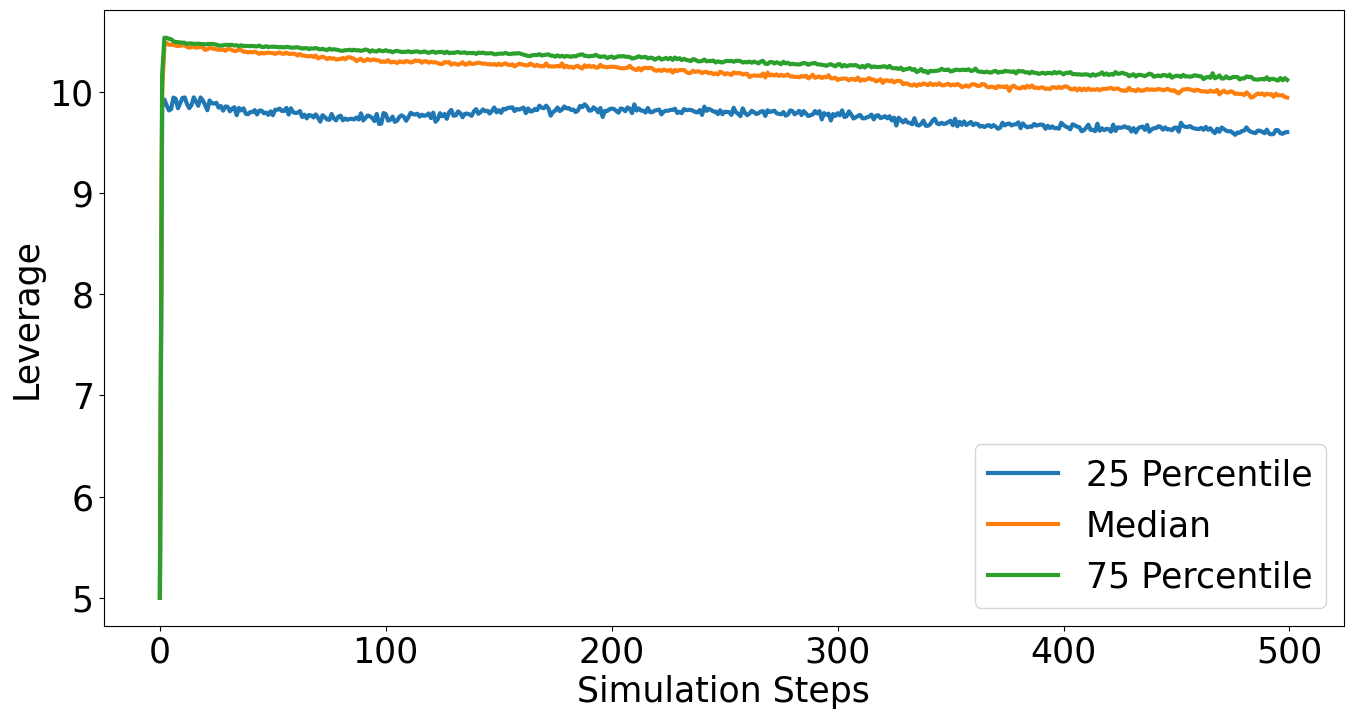}
         \caption{Leverage Ratio Distribution}
         \label{leverageRaio}
    \end{subfigure}
    \hfill
    \caption{System Statistics}
    \label{systemStatistics} 
\end{figure}

Figure~\ref{inOutDegree} compares the degree distribution of the simulated network to the empirical distribution. The model reproduces the skewed distribution of connectivity, with a concentration of links among core institutions. Figure~\ref{systemStatistics} further examines exposure patterns and centrality measures. The simulated network captures the concentration of interbank exposures and the dominance of central nodes, consistent with observed network characteristics.

\begin{figure}[ht]
    \centering
    \begin{subfigure}[b]{0.495\textwidth}
        \centering    
        \includegraphics[width=\textwidth]{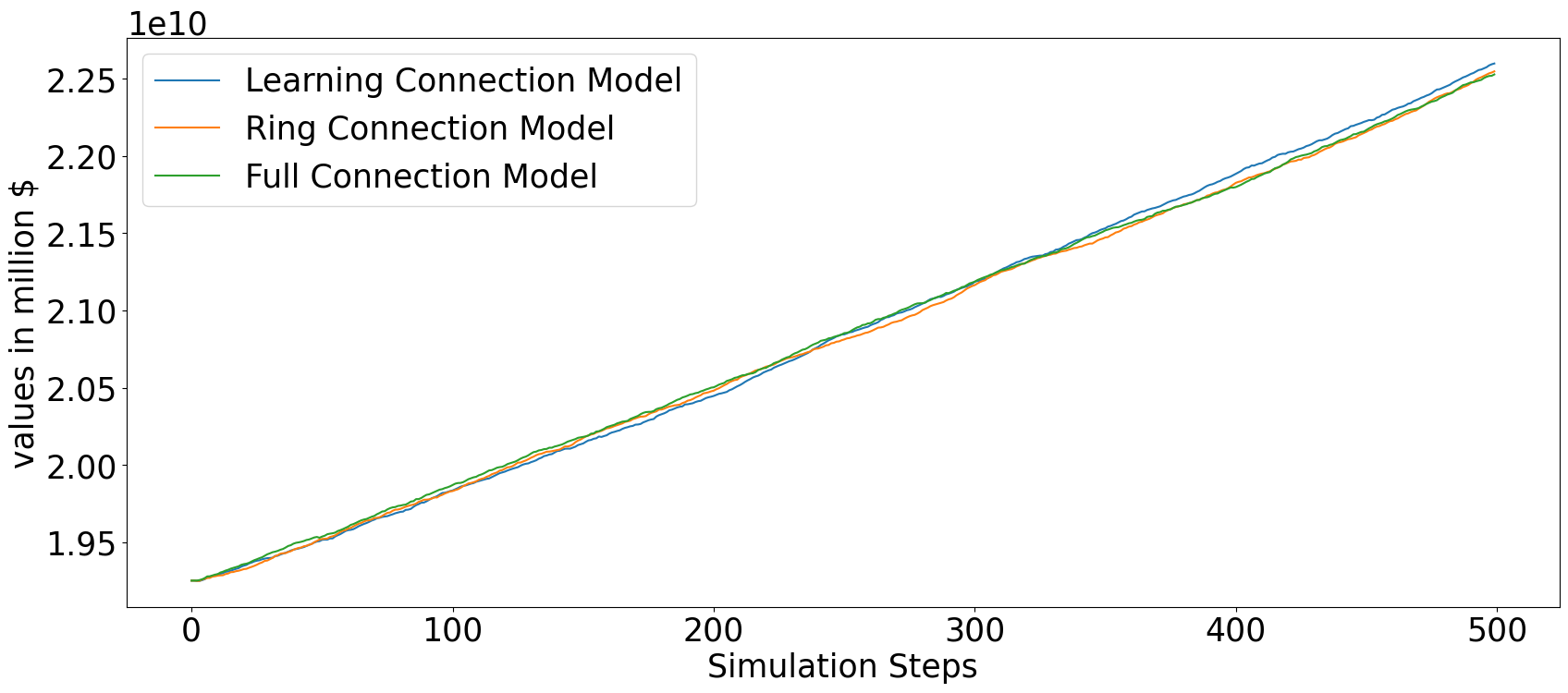}
         \caption{Intra-Financial System Assets}
         \label{IntraAssets}
    \end{subfigure}
    \hfill
    \begin{subfigure}[b]{0.495\textwidth}        
    \centering
        \includegraphics[width=\textwidth]{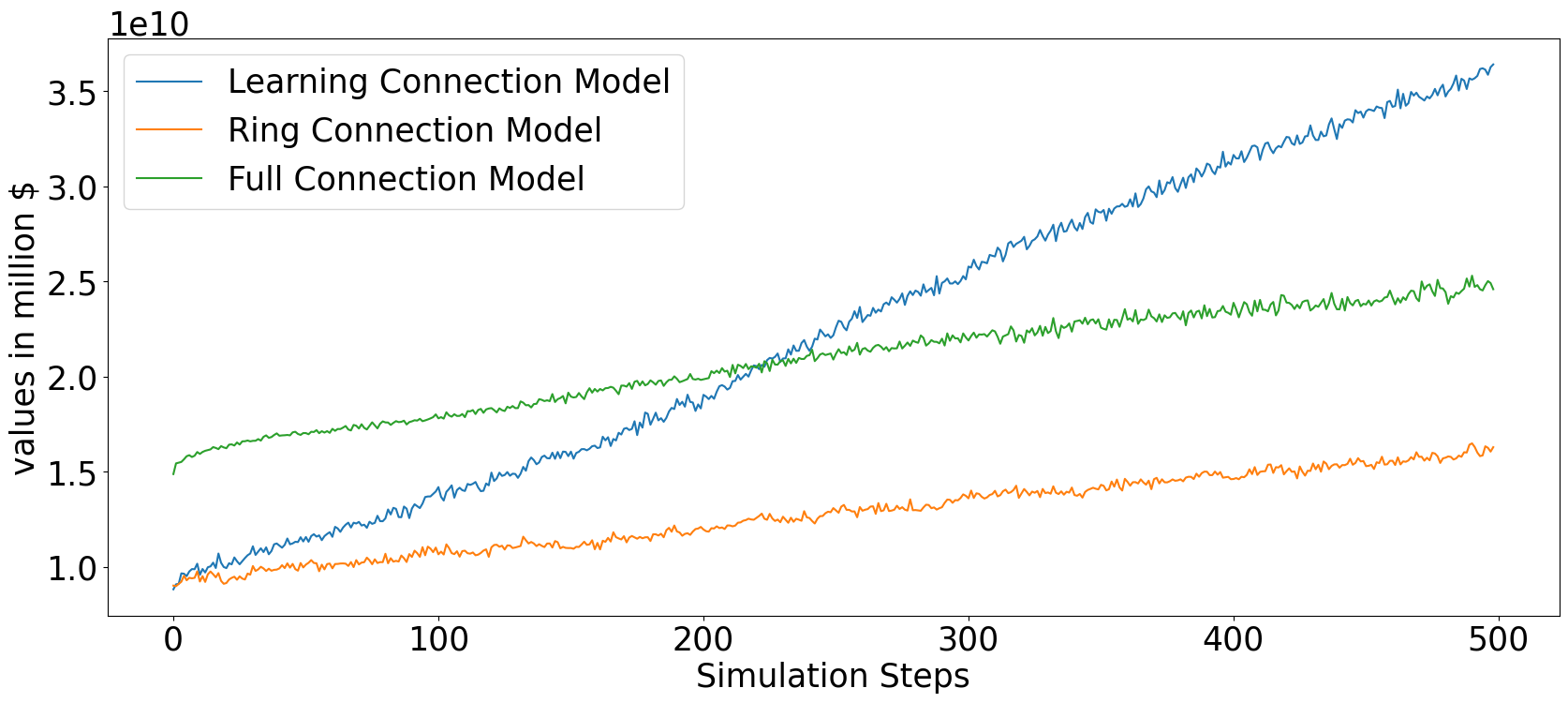}
         \caption{Intra-Financial System Liabilities}
         \label{IntraLiabilities}
    \end{subfigure}
    \hfill
    \caption{Intra-Financial System Assets and Liabilities}
    \label{intraFinancialSystem} 
\end{figure}

Finally, Figure~\ref{intraFinancialSystem} presents the distribution of losses and default frequencies under baseline conditions. To demonstrate that the network structure matters, we compare the network structure with two of the highly stylized network structures: the ring-shape network and the full-connect network. We define the ring-shape relationship connection where each bank borrows only from the next slightly larger bank; the structure forms a unidirectional loop of credit relationships that goes from the smallest to the largest banks. We define the full-connect network structure as one where each bank has direct lending and borrowing relationships with every other bank, and the chances of a bank borrowing from other banks are equally likely. The learning connection model shows a substantial and consistent liability increase over time. This implies the adaptive nature of the learning model, where banks gradually form stable lending and borrowing relations due to optimizing the efficiency of lending and borrowing. The model generates clustered defaults and nonlinear responses to shocks, reflecting the amplification effects of correlated exposures and network structure.

\subsection{Discussion}

Overall, the calibrated model replicates key empirical features of the banking system, including asset distribution, leverage, and network concentration. The close alignment between simulated and observed moments supports the use of the model as a framework to analyze how accounting regimes influence the buildup and propagation of systemic risk.

\section{Main Results}
\label{sec:results}

This section presents the main results of the model, focusing on how asset return correlation, the timing of loss recognition, and interbank network structure jointly determine systemic risk. We begin by describing the experimental design and parameter configurations used in the simulations, and then present the main findings.

\subsection{Experimental Design}

To quantify the effects of correlation, accounting regimes, and network interactions, we conduct a series of simulation experiments under different parameter configurations. Table~\ref{systemicRisk} summarizes the key parameters used in the experiments, including the number of banks, shock characteristics, and accounting assumptions.

\begin{table}[ht]
\centering
\resizebox{0.8 \columnwidth}{!}{%
\begin{tabular}{c c c c c c c c} 
\bb{Correlation Effect}  & Base Model & High Correlation & No Correlation\\
 \hline
Leverage Ratio Upper Bound $\overline{LR}$   & 10.0  & 10.0 & 10.0 \\
Capital Reserve as a ratio of Deposit        & 0.20  & 0.20 &  0.20 \\
Volatility Spike $\Delta_{\bo{\sigma}}$      & 5.00  & 5.00 & 5.00 \\
Portfolio Returns Correlation $\rho$         & 0.80 & \bb{\R{0.90}} & \bb{\R{0.00}} \\
Number of borrowing requests                 & 5    & 5   & 5  \\
Time to receive shock $t \in [t_{\text{begin}}, t_{\text{end}}]$ & [300,300]  & [300,300] & [300,300] \\
\hline
\\
\end{tabular}
}%

\resizebox{0.8 \columnwidth}{!}{%
\begin{tabular}{c c c c c c c c} 
\bb{One Shock vs Multiple Shocks}  & Base Model & 2 Shocks & 3 Shocks\\
 \hline
Leverage Ratio Upper Bound $\overline{LR}$   & 10.0  & 10.0 & 10.0 \\
Capital Reserve as a ratio of Deposit        & 0.20   & 0.20 & 0.20 \\
Volatility Spike $\Delta_{\bo{\sigma}}$      & 5.00   & 5.00 & 5.00 \\
Portfolio Returns Correlation $\rho$         & 0.80  & 0.80 & 0.80 \\
Number of borrowing requests                 & 5      & 5  & 5\\
Time to receive shock $t \in [t_{\text{begin}}, t_{\text{end}}]$  & [300,300] &  \bb{\R{[299,300]}}  &  \bb{\R{[298,300]}} \\
\hline
\\
\end{tabular}
}%

\caption{Parameter of the Experiments}
\label{systemicRisk}
\end{table}

The baseline setup consists of $N = 100$ banks with balance sheet characteristics calibrated as described in Section 5. We vary two key dimensions. First, we vary the correlation parameter $\rho$ to capture different degrees of common exposure to interest rate risk. Second, we vary the timing of loss realization by comparing scenarios with gradual loss recognition (MTM) and concentrated loss realization (HTM).

\begin{figure}[ht]
     \centering
     \includegraphics[width=0.8\textwidth]{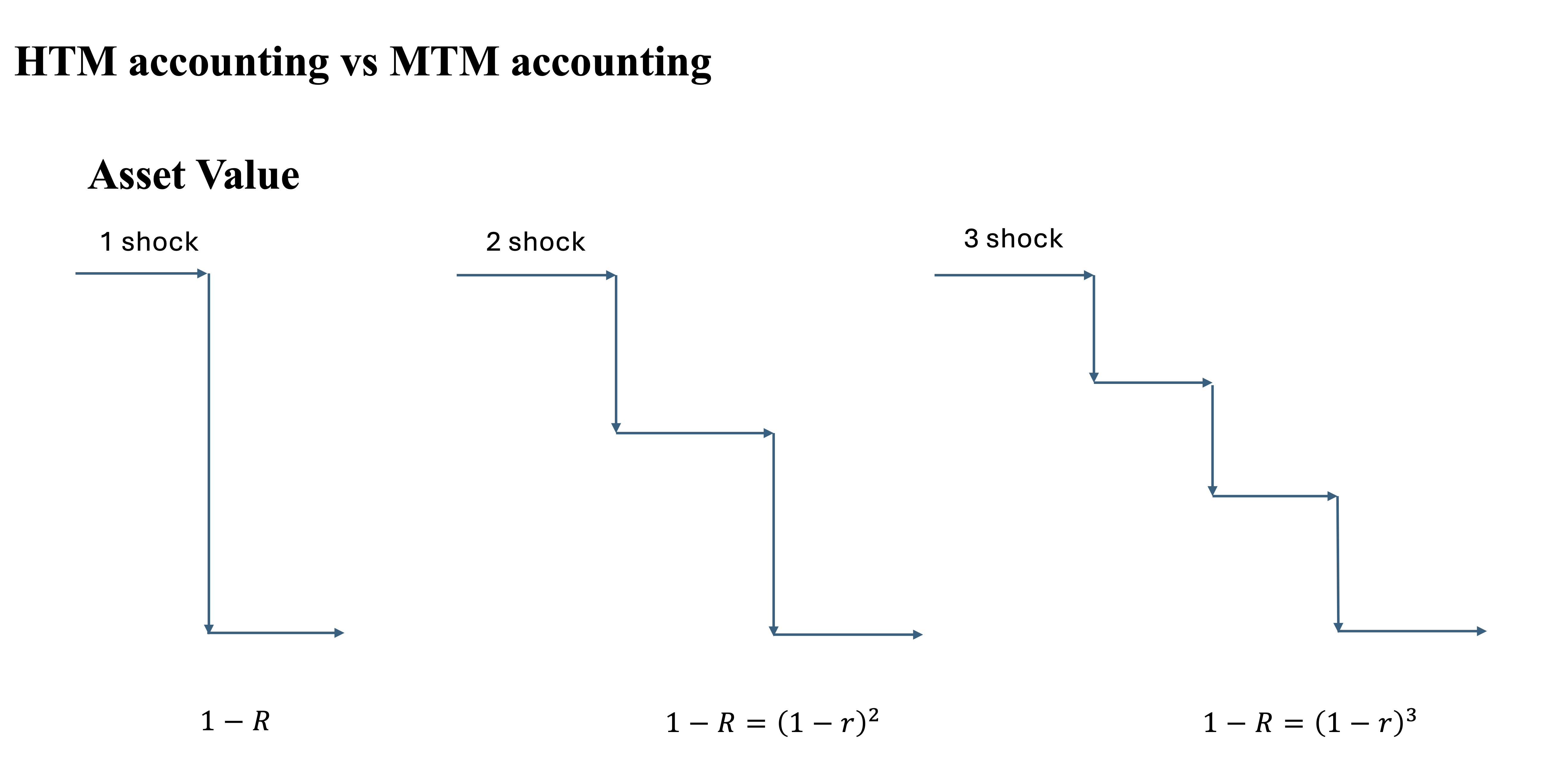}
     \caption{Essential difference in the asset value between HTM and MTM}
     \label{HTMvsMTM}
\end{figure}

In addition, we consider different shock structures, including single-period shocks and multi-period shocks that distribute losses over time (see Figure~\ref{HTMvsMTM}). By doing this, the system receives the same accumulated asset shock in all three settings (one single shock, two shocks, and three shocks). But when we replace the HTM accounting with the fair-value accounting, the shock will be reflected in the bank balance sheet in a more up-to-date fashion and reveal the losses in the banks’ balance sheet immediately instead of accumulating the losses and revealing them altogether. The change in the asset values of banks will be more gradual, reducing the volatility in the bank's financial reports, and making the masking of the real-time risk of the assets impossible. This design allows us to isolate the effects of correlation and timing while keeping the total magnitude of losses constant.

The outcome variable of interest is the number of bank defaults following a shock, which provides a direct measure of systemic risk. All results are averaged across multiple simulation runs to ensure robustness.

\subsection{Correlation and Nonlinear Systemic Risk}

We begin by examining the role of asset return correlation in shaping systemic outcomes. Table~\ref{defaultNumber_corr} reports the average number of bank defaults under different levels of correlation $\rho$. The results reveal a highly nonlinear relationship between correlation and systemic risk.

\begin{table}[ht]
\centering
\resizebox{0.8 \columnwidth}{!}{%
    \begin{tabular}{lrrrrr}
    \hline
    {} &  Base Model &  High Correlation &  No Correlation &  2 Shock &  3 Shock \\
    \hline
    mean  &       12.03 &            58.37 &           11.73 &     1.68 &     0.08 \\
    std   &       20.01 &            33.79 &            3.20 &     7.98 &     1.51 \\
    min   &        0.00 &             0.00 &            3.00 &     0.00 &     0.00 \\
    25\%   &        0.00 &             27.00 &            9.00 &     0.00 &     0.00 \\
    50\%   &        2.00 &             55.00 &           12.00 &     0.00 &     0.00 \\
    75\%   &       14.00 &            98.00 &           14.00 &     0.00 &     0.00 \\
    max   &      80.00 &            100.00 &           22.00 &    99.00 &    55.00 \\
    \hline
    \end{tabular}
}%
\caption{Number of Defaults over 2000 Simulation}
\label{defaultNumber_corr}
\end{table}

As $\rho$ increases from moderate to high levels, the number of defaults rises disproportionately. In particular, increasing correlation from 0.8 to 0.9 leads to a sharp escalation in defaults, with the average number of failed banks increasing from approximately 12 to over 58. This nonlinear response reflects the increasing likelihood of simultaneous balance sheet deterioration across institutions as asset returns become more synchronized.

Figure~\ref{distributionDefault}a illustrates this pattern graphically. At lower levels of correlation, shocks are relatively dispersed across banks, allowing the system to absorb losses through diversification. However, as correlation rises, losses become increasingly synchronized, reducing the effectiveness of diversification and leading to clustered defaults.

These findings are consistent with the mechanism described in Section 3. When banks hold similar interest rate–sensitive assets, increases in interest rates generate highly correlated losses, making systemic risk particularly sensitive to changes in correlation.

\subsection{Timing of Loss Recognition: HTM vs MTM}

We next examine how accounting regimes affect systemic outcomes by altering the timing of loss recognition.

\begin{figure}[ht]
     \centering
     \begin{subfigure}[b]{\textwidth}
         \centering
         \includegraphics[width=\textwidth]{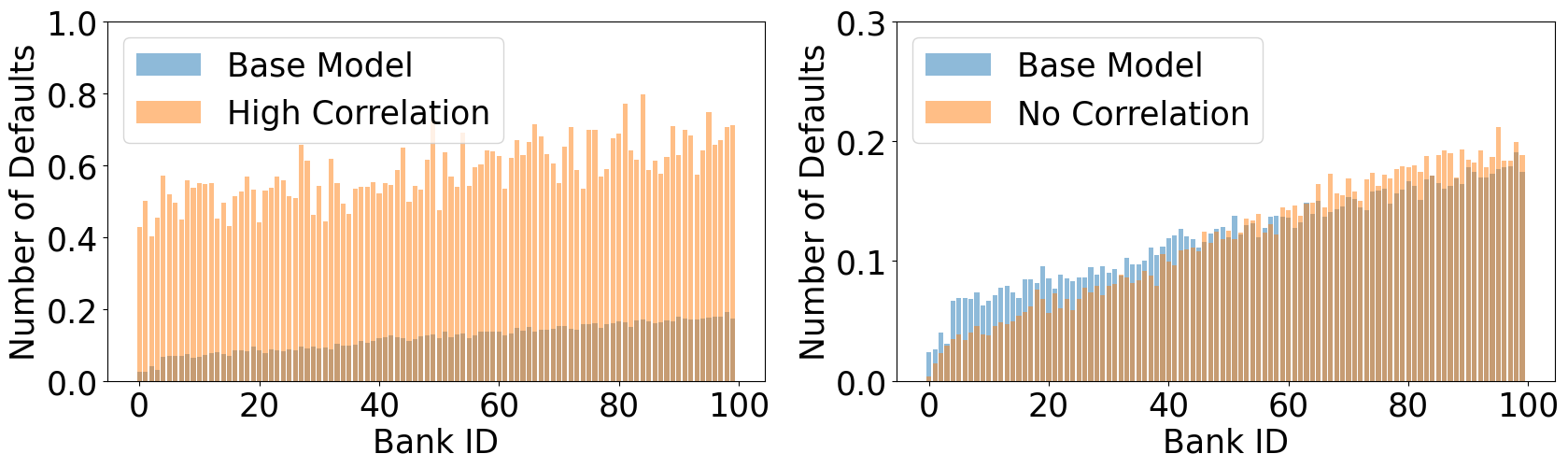}
         \caption{High Correlation and No Correlation Between Asset Shocks}
         \label{fig: lowNoCorrelation}
     \end{subfigure}
     \begin{subfigure}[b]{\textwidth}
         \centering
         \includegraphics[width=\textwidth]{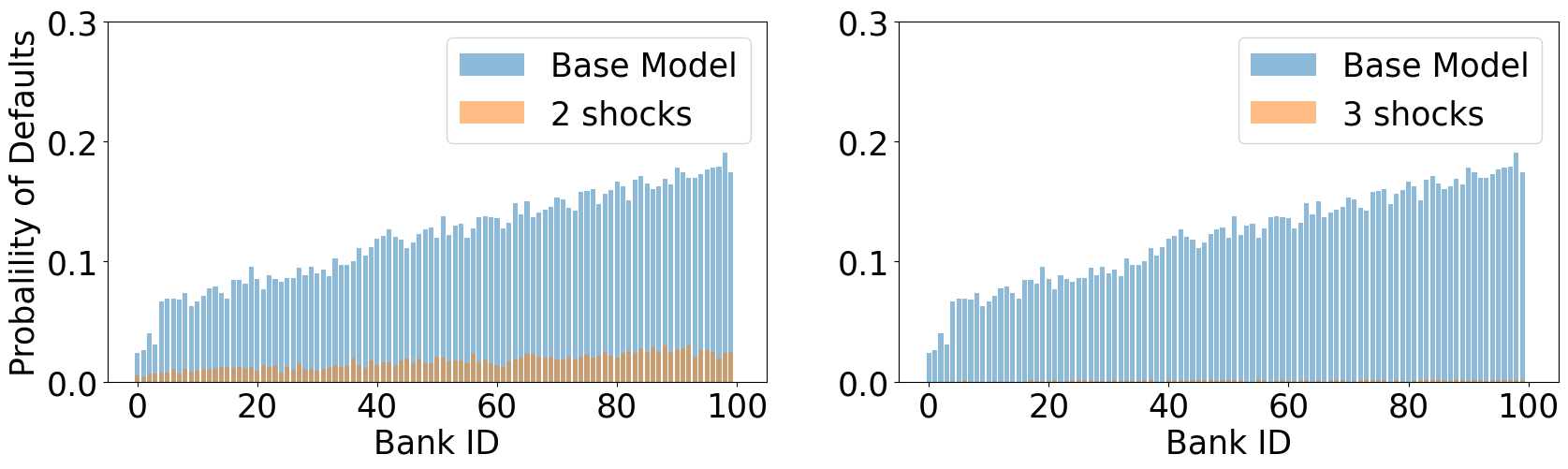}
         \caption{Single Shock vs Multiple Shocks}
         \label{multipleShocks}
     \end{subfigure}
     \caption{Probability of Defaults During Shock Simulation  \\ In this figure, Bank ID is ordered based on the asset size of the bank where 0 is the bank with the largest asset holding and 99 is the bank with the smallest asset holding.}
    \label{distributionDefault}
\end{figure}

Figure~\ref{distributionDefault} provides a conceptual illustration of this mechanism. Under MTM accounting, losses are recognized continuously as asset values decline, allowing banks to adjust their balance sheets over time. In contrast, under HTM accounting, losses accumulate without affecting reported equity until assets are liquidated, at which point losses are realized abruptly.

The quantitative implications are shown in Table~\ref{defaultNumber_corr}. Holding total losses constant, concentrated loss realization leads to significantly higher default rates. When losses are realized in a single period, the average number of defaults is approximately 12.03. Distributing the same losses across two periods reduces defaults to 1.68, while further smoothing across three periods reduces defaults to 0.08.

Figure~\ref{multipleShocksMonitor} reinforces this result by showing the distribution of defaults under different timing scenarios. Gradual loss recognition allows banks to adjust leverage and exposures, reducing systemic stress. In contrast, concentrated losses trigger abrupt balance sheet contractions and widespread defaults.

These results demonstrate that the timing of loss recognition is a key determinant of systemic risk. Delayed recognition can amplify instability by synchronizing losses across banks, even when aggregate losses are unchanged.

\subsection{Network Amplification and Contagion}

Finally, we examine how interbank network structure amplifies shocks.

Figure~\ref{distributionDefault}b shows the distribution of defaults across banks in the network. Defaults are concentrated among highly connected institutions, which play a central role in interbank lending. When these core banks fail, the impact on the system is disproportionately large.

The network amplifies shocks through interbank exposures. When a borrower defaults, its creditors incur losses, which may trigger further defaults. This process generates cascade effects that depend on both the size of exposures and the network topology.

The interaction between correlation, timing, and network structure is central. High correlation increases the likelihood of simultaneous distress, delayed loss recognition synchronizes shock realization, and network linkages propagate these shocks across institutions. Together, these mechanisms produce nonlinear increases in systemic risk.

\subsection{Summary of Findings}

The results highlight three key insights. First, systemic risk increases nonlinearly with asset return correlation. Second, the timing of loss recognition significantly affects system stability, with delayed recognition amplifying systemic risk. Third, interbank networks act as amplification mechanisms, transforming correlated and synchronized shocks into widespread financial instability.

These findings underscore the importance of accounting regimes and network structure in shaping the resilience of the banking system.

\section{Policy Implications}
\label{sec:implications}

The results of the model have several important implications for financial regulation, accounting design, and systemic risk monitoring. By highlighting the interaction between accounting-induced opacity, correlated asset exposures, and interbank network structure, the analysis suggests that policies affecting the timing and transparency of loss recognition can have significant effects on financial stability.

\subsection{Accounting Design and Loss Recognition}

A central implication of the analysis is that accounting rules governing the timing of loss recognition play a critical role in shaping systemic risk. While HTM accounting can reduce short-term volatility in reported earnings and regulatory capital, the model shows that it may also facilitate the accumulation of unrealized losses that are realized abruptly under stress.

\begin{figure}[ht]
     \centering
     \includegraphics[width=0.6\textwidth]{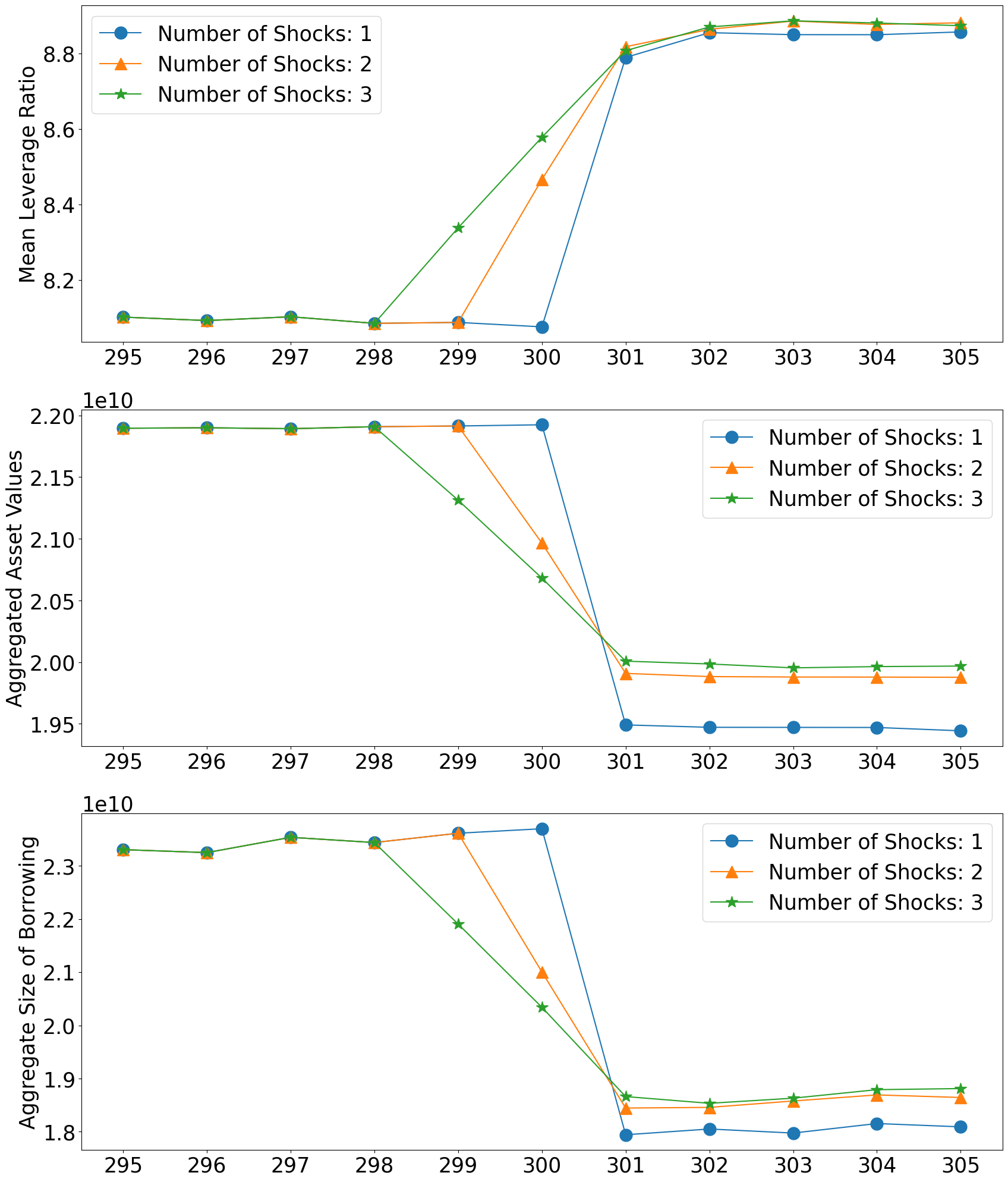}
     \caption{Single Shock vs Multiple Shocks $t \in [295, 305]$}
     \label{multipleShocksMonitor}
\end{figure}

This result is directly reflected in Table~\ref{defaultNumber_corr} and Figure~\ref{multipleShocksMonitor}, which show that concentrated loss realization leads to significantly higher default rates than gradual loss recognition, even when total losses are held constant. These findings are consistent with the broader accounting literature emphasizing that financial reporting choices affect the visibility of bank risk and the interpretation of bank solvency \citep{BeattyLiao2014, laux2010did, huizinga2012bank}. More recent work further suggests that delayed loss recognition can increase fragility by allowing risks to build up over time \citep{BegenauBigioMajerovitzVieyra2024}.

These results suggest that accounting regimes should be evaluated not only in terms of their impact on individual bank reporting but also in terms of their systemic consequences. In particular, delaying loss recognition may improve the appearance of stability in normal times while increasing vulnerability to sudden and widespread disruptions.

One implication is that regulators may need to incorporate additional measures of economic value into supervisory frameworks. Enhanced disclosure of unrealized losses or supplementary mark-to-market metrics could reduce the gap between reported and economic balance sheet conditions, allowing both regulators and market participants to better assess underlying risks.

\subsection{Interaction with Monetary Policy}

The results also highlight an important interaction between accounting regimes and monetary policy. Periods of monetary tightening, particularly those involving rapid increases in interest rates, can generate large valuation losses on long-duration assets. As shown in Table~\ref{defaultNumber_corr} and Figure~\ref{distributionDefault}a, when these losses are highly correlated across banks, systemic risk increases nonlinearly.

This finding is consistent with recent empirical evidence documenting that unrealized losses on bank balance sheets became systemically important during the recent tightening cycle \citep{jiang2024monetary}. When losses are not immediately recognized under HTM accounting, they may accumulate across institutions and remain hidden until liquidity pressures force asset sales.

These results suggest that the effects of monetary policy on financial stability depend not only on macroeconomic conditions but also on accounting and regulatory frameworks. In particular, tightening cycles may have stronger destabilizing effects when banks hold large HTM portfolios, as losses remain unrecognized until they are realized in a synchronized manner.

From a policy perspective, this highlights the importance of monitoring interest rate risk and unrealized losses across the banking system, particularly during periods of rapid monetary tightening.

\subsection{Network Structure and Systemic Risk Monitoring}

The model further emphasizes the role of interbank networks in amplifying systemic risk. As shown in Figure~\ref{distributionDefault}b, defaults are concentrated among highly connected institutions, whose failure generates disproportionate spillovers to the rest of the system.

This result is consistent with the financial network literature, which shows that interconnectedness can amplify shocks and generate cascade effects \citep{eisenberg2001systemic, glasserman2016contagion, CabralesGottardiVegaRedondo2017}. More recent work also highlights the importance of network centrality in determining systemic importance \citep{DenbeeJulliardLiYuan2021}.

In addition, the interaction between network structure and accounting opacity introduces a new dimension of systemic risk. Because HTM accounting affects reported equity, it can influence perceived bank strength and, in turn, interbank exposure allocation. As a result, banks with hidden losses may become more central in the network, increasing the potential for systemic spillovers when losses are realized.

These findings suggest that systemic risk monitoring should account for both balance sheet exposures and network structure. Measures of systemic importance should incorporate not only size and leverage but also interconnectedness and centrality.

\subsection{Implications for Stress Testing and Regulation}

The findings also have implications for stress testing frameworks. Standard stress tests often assume that losses are realized gradually and independently across institutions. However, the model shows that both the timing and correlation of losses are critical determinants of systemic outcomes.

As illustrated in Table~\ref{defaultNumber_corr} and Figures~\ref{distributionDefault}, scenarios with correlated shocks and synchronized loss realization produce significantly higher levels of systemic risk. This suggests that stress testing frameworks should incorporate scenarios that account for both correlated exposures and alternative accounting treatments.

In particular, incorporating HTM-style delayed loss recognition into stress testing exercises may provide a more realistic assessment of systemic vulnerabilities. This is especially relevant in environments where banks hold large portfolios of interest rate–sensitive securities.

More broadly, the results suggest that regulatory frameworks should consider the interaction between accounting rules, asset composition, and network structure. Policies that improve transparency, limit excessive concentration in common exposures, or reduce interconnectedness may help mitigate systemic risk.

Overall, the analysis highlights that accounting regimes are not neutral from a systemic perspective. By affecting the timing and visibility of losses, accounting rules influence both the buildup of risk and its propagation through the financial system. These findings suggest that financial stability considerations should play a central role in the design of accounting standards and regulatory policies.

\section{Conclusion}
\label{sec:conclusion}
This paper studies how accounting-induced opacity interacts with asset commonality and financial interconnectedness to shape systemic risk in the banking system. Motivated by the rapid increase in interest rates during recent monetary tightening episodes and the associated accumulation of unrealized losses on bank balance sheets, we develop a framework that links accounting treatment, asset correlation, and network structure to financial stability.

The analysis highlights a mechanism through which HTM accounting can contribute to systemic fragility. By delaying the recognition of valuation losses, the HTM accounting allows losses to accumulate over time without affecting the reported capital. When banks hold similar interest rate--sensitive assets, these losses are highly correlated across institutions. If liquidity pressures force asset sales, the realization of these losses can occur in a synchronized manner, leading to abrupt balance sheet adjustments across the system. Interbank connections further amplify these shocks, generating cascade effects that increase systemic risk.

Using a calibrated model with endogenous network formation, we show that systemic risk depends critically on both the correlation of asset returns and the timing of loss recognition. Small increases in asset return correlation can lead to disproportionately large increases in bank defaults. Moreover, concentrating losses in a single period, as implied by delayed recognition under HTM accounting, significantly amplifies systemic risk relative to gradual loss realization. These effects are further magnified by the structure of the interbank network, which transmits shocks across institutions.

The results contribute to a growing literature on financial stability by emphasizing that accounting rules are not neutral from a systemic perspective. By affecting the timing and visibility of losses, accounting regimes influence not only individual bank outcomes but also the propagation of shocks through the financial system. These findings suggest that evaluating accounting standards requires consideration of their macroprudential implications, particularly in environments characterized by common asset exposures and strong financial linkages.

Several directions for future research emerge from this analysis. First, incorporating more detailed institutional features, such as heterogeneous depositor behavior or regulatory constraints, may provide additional insights into the interaction between liquidity risk and accounting treatment. Second, extending the framework to incorporate multiple asset classes and endogenous portfolio choice could further illuminate the role of asset allocation in systemic risk. Finally, integrating the model with empirical measures of unrealized losses and network exposures may help improve the design of stress testing and macroprudential policy.

Overall, the paper highlights the importance of considering accounting, market structure, and financial networks jointly when assessing systemic risk. As financial systems continue to evolve, understanding these interactions will remain central to the design of policies aimed at promoting financial stability.

\bibliographystyle{model5-names}
\biboptions{authoryear}
\bibliography{references.bib}

@article{granja2023bank,
  title={Bank fragility and reclassification of securities into HTM},
  author={Granja, Joao},
  journal={Available at SSRN 4409834},
  year={2023}
}

@article{glasserman2016contagion,
  title={Contagion in financial networks},
  author={Glasserman, Paul and Young, H Peyton},
  journal={Journal of Economic Literature},
  volume={54},
  number={3},
  pages={779--831},
  year={2016},
  publisher={American Economic Association 2014 Broadway, Suite 305, Nashville, TN 37203-2425}
}

@article{cifuentes2005liquidity,
  title={Liquidity risk and contagion},
  author={Cifuentes, Rodrigo and Ferrucci, Gianluigi and Shin, Hyun Song},
  journal={Journal of the European Economic association},
  volume={3},
  number={2-3},
  pages={556--566},
  year={2005},
  publisher={Oxford University Press}
}

@article{caballero2013fire,
  title={Fire sales in a model of complexity},
  author={Caballero, Ricardo J and Simsek, Alp},
  journal={The Journal of Finance},
  volume={68},
  number={6},
  pages={2549--2587},
  year={2013},
  publisher={Wiley Online Library}
}

@article{tasca2016market,
  title={Market procyclicality and systemic risk},
  author={Tasca, Paolo and Battiston, Stefano},
  journal={Quantitative Finance},
  volume={16},
  number={8},
  pages={1219--1235},
  year={2016},
  publisher={Taylor \& Francis}
}

@article{bookstaber2018agent,
  title={An agent-based model for financial vulnerability},
  author={Bookstaber, Richard and Paddrik, Mark and Tivnan, Brian},
  journal={Journal of Economic Interaction and Coordination},
  volume={13},
  number={2},
  pages={433--466},
  year={2018},
  publisher={Springer}
}

@article{eisenberg2001systemic,
  title={Systemic risk in financial systems},
  author={Eisenberg, Larry and Noe, Thomas H},
  journal={Management Science},
  volume={47},
  number={2},
  pages={236--249},
  year={2001},
  publisher={INFORMS}
}

@article{BeattyLiao2014,
  author = {Beatty, Anne and Liao, Scott},
  title = {Financial Accounting in the Banking Industry: A Review of the Empirical Literature},
  journal = {Journal of Accounting and Economics},
  year = {2014},
  volume = {58},
  pages = {339--383}
}

@article{BegenauBigioMajerovitzVieyra2024,
  author = {Begenau, Juliane and Bigio, Saki and Majerovitz, Julia and Vieyra, Manuel},
  title = {Delayed Loss Accounting and Banking Fragility},
  journal = {Review of Economic Studies},
  year = {2024}
}

@article{jiang2024monetary,
  title={Monetary tightening and US bank fragility in 2023: Mark-to-market losses and uninsured depositor runs?},
  author={Jiang, Erica Xuewei and Matvos, Gregor and Piskorski, Tomasz and Seru, Amit},
  journal={Journal of Financial Economics},
  volume={159},
  pages={103899},
  year={2024},
  publisher={Elsevier}
}

@article{GranjaEtAl2024,
  author = {Granja, Joao and Jiang, Erica and Matvos, Gregor and Piskorski, Tomasz and Seru, Amit},
  title = {Book Value Risk Management of Banks: Limited Hedging, HTM Accounting, and Rising Interest Rates},
  journal = {Working Paper},
  year = {2024}
}

@article{Jungherr2018,
  author = {Jungherr, Jens},
  title = {Bank Opacity and Financial Crises},
  journal = {Journal of Banking \& Finance},
  year = {2018},
  volume = {97},
  pages = {157--176}
}

@article{AcharyaThakor2016,
  author = {Acharya, Viral V. and Thakor, Anjan},
  title = {The Dark Side of Liquidity Creation},
  journal = {Journal of Financial Intermediation},
  year = {2016},
  volume = {28},
  pages = {4--21}
}

@article{RogersVeraart2013,
  author = {Rogers, Leonard and Veraart, Luitgard},
  title = {Failure and Rescue in an Interbank Network},
  journal = {Management Science},
  year = {2013},
  volume = {59},
  pages = {882--898}
}

@article{DenbeeJulliardLiYuan2021,
  author = {Denbee, Edward and Julliard, Christian and Li, Ye and Yuan, Kathy},
  title = {Network Risk and Key Players: A Structural Analysis of Interbank Liquidity},
  journal = {Journal of Financial Economics},
  year = {2021},
  volume = {141},
  pages = {831--859}
}

@article{CabralesGottardiVegaRedondo2017,
  author = {Cabrales, Antonio and Gottardi, Piero and Vega-Redondo, Fernando},
  title = {Risk Sharing and Contagion in Networks},
  journal = {Review of Financial Studies},
  year = {2017},
  volume = {30},
  pages = {3086--3127}
}

@article{ChoiGoldsmithPinkhamYorulmazer2023,
  author = {Choi, David and Goldsmith-Pinkham, Paul and Yorulmazer, Tanju},
  title = {Contagion Effects of the Silicon Valley Bank Run},
  journal = {Working Paper},
  year = {2023}
}

@article{greenwood2015vulnerable,
  title={Vulnerable banks},
  author={Greenwood, Robin and Landier, Augustin and Thesmar, David},
  journal={Journal of financial economics},
  volume={115},
  number={3},
  pages={471--485},
  year={2015},
  publisher={Elsevier}
}

@article{liu_interbank_2020,
	title = {Interbank contagion: {An} agent-based model approach to endogenously formed networks},
	volume = {112},
	issn = {03784266},
	shorttitle = {Interbank contagion},
	url = {https://linkinghub.elsevier.com/retrieve/pii/S0378426617301942},
	doi = {10.1016/j.jbankfin.2017.08.008},
	language = {en},
	urldate = {2023-02-05},
	journal = {Journal of Banking \& Finance},
	author = {Liu, Anqi and Paddrik, Mark and Yang, Steve Y. and Zhang, Xingjia},
	month = mar,
	year = {2020},
	pages = {105191},
}

@article{lux_emergence_2015,
	title = {Emergence of a core-periphery structure in a simple dynamic model of the interbank market},
	volume = {52},
	issn = {01651889},
	url = {https://linkinghub.elsevier.com/retrieve/pii/S0165188914002607},
	doi = {10.1016/j.jedc.2014.09.038},
	language = {en},
	urldate = {2023-02-05},
	journal = {Journal of Economic Dynamics and Control},
	author = {Lux, Thomas},
	month = mar,
	year = {2015},
	pages = {A11--A23},
}

@article{elliott_systemic_2021,
	title = {Systemic risk shifting in financial networks},
	volume = {191},
	issn = {00220531},
	url = {https://linkinghub.elsevier.com/retrieve/pii/S0022053120301502},
	doi = {10.1016/j.jet.2020.105157},
	language = {en},
	urldate = {2023-02-05},
	journal = {Journal of Economic Theory},
	author = {Elliott, Matthew and Georg, Co-Pierre and Hazell, Jonathon},
	month = jan,
	year = {2021},
	pages = {105157},
}

@article{anderson2019bank,
  title={Bank networks and systemic risk: Evidence from the national banking acts},
  author={Anderson, Haelim and Paddrik, Mark and Wang, Jessie Jiaxu},
  journal={American Economic Review},
  volume={109},
  number={9},
  pages={3125--3161},
  year={2019},
  publisher={American Economic Association 2014 Broadway, Suite 305, Nashville, TN 37203}
}

@article{houy2020defaulting,
  title={Defaulting firms and systemic risks in financial networks: a normative approach},
  author={Houy, Nicolas and Jouneau, Fr{\'e}d{\'e}ric and Le Grand, Fran{\c{c}}ois},
  journal={Economic Theory},
  pages={503--526},
  year={2020},
  publisher={JSTOR}
}

@misc{jackson_systemic_2020,
	title = {Systemic {Risk} in {Financial} {Networks}: {A} {Survey}},
	shorttitle = {Systemic {Risk} in {Financial} {Networks}},
	url = {http://arxiv.org/abs/2012.12702},
	language = {en},
	urldate = {2023-02-06},
	publisher = {arXiv},
	author = {Jackson, Matthew O. and Pernoud, Agathe},
	month = dec,
	year = {2020},
	note = {Number: arXiv:2012.12702
arXiv:2012.12702 [physics, q-fin]},
        howpublished = {}
}

@article{duarte2021fire,
  title={Fire-sale spillovers and systemic risk},
  author={Duarte, Fernando and Eisenbach, Thomas M},
  journal={The Journal of Finance},
  volume={76},
  number={3},
  pages={1251--1294},
  year={2021},
  publisher={Wiley Online Library}
}

@article{soramaki2007topology,
  title={The topology of interbank payment flows},
  author={Soram{\"a}ki, Kimmo and Bech, Morten L and Arnold, Jeffrey and Glass, Robert J and Beyeler, Walter E},
  journal={Physica A: Statistical Mechanics and its Applications},
  volume={379},
  number={1},
  pages={317--333},
  year={2007},
  publisher={Elsevier}
}

@article{duffie2018financial,
  title={Financial regulatory reform after the crisis: An assessment},
  author={Duffie, Darrell},
  journal={Management Science},
  volume={64},
  number={10},
  pages={4835--4857},
  year={2018},
  publisher={INFORMS}
}

@article{demirgucc2021banking,
  title={Banking sector performance during the COVID-19 crisis},
  author={Demirg{\"u}{\c{c}}-Kunt, Asli and Pedraza, Alvaro and Ruiz-Ortega, Claudia},
  journal={Journal of Banking \& Finance},
  volume={133},
  pages={106305},
  year={2021},
  publisher={Elsevier}
}

@article{huizinga2012bank,
  title={Bank valuation and accounting discretion during a financial crisis},
  author={Huizinga, Harry and Laeven, Luc},
  journal={Journal of financial economics},
  volume={106},
  number={3},
  pages={614--634},
  year={2012},
  publisher={Elsevier}
}

@article{laux2010did,
  title={Did fair-value accounting contribute to the financial crisis?},
  author={Laux, Christian and Leuz, Christian},
  journal={Journal of economic perspectives},
  volume={24},
  number={1},
  pages={93--118},
  year={2010},
  publisher={American Economic Association}
}

@article{begenau2026q,
  title={A q-theory of banks},
  author={Begenau, Juliane and Bigio, Saki and Majerovitz, Jeremy and Vieyra, Matias},
  journal={Review of Economic Studies},
  volume={93},
  number={1},
  pages={106--143},
  year={2026},
  publisher={Oxford University Press UK}
}

\section*{Appendix}
\label{appendix}

\subsection{HTM Holdings vs. Asset Correlation}
\label{app:asset_correlation}
To test the idea that increased HTM holdings suggest higher asset return correlations, we come up with the following definition of Asset-Weighted Correlation. In a typical correlation matrix of asset returns, each element measures the correlation between two banks' assets. The correlation helps understand pairwise relationships, but it is not easy to understand the systemwide level of the return correlation. An asset-weighted correlation could address this issue by weighting each correlation value by the product of the asset weights of the two banks and summing up all the pairwise asset-weighted correlations to get a single number. This single number helps us get an idea of the aggregated correlation level among a large number of banks. Based on the asset size of each bank, we could calculate the asset weight of banks, $w_i =  \frac{E_i}{\sum_{j=1}^{n} E_j}$, given the pairwise asset return correlation $\rho_{i,j}$, the asset-weighted correlation is defined as: 

$$
    \text{AW} = \sum_{i<j} w_i \cdot w_j \cdot \rho_{ij}
$$

Based on the market data of the 20 largest public-traded banks in the U.S. from 2004 to 2023, we observed the elevated asset-weighted correlation in recent years and the sharp increase in the asset-weighted HTM Ratio. We conduct two simple linear regression analyses, regressing the asset-weighted correlations on the asset-weighted Securities Ratio and regressing the asset-weighted correlations on the asset-weighted HTM Ratio. The results in Table \ref{regressionResults} show that the Securities Ratio and HTM Ratio both positively predict the aggregated correlation level among banks.

\begin{table}[ht]
\centering
    \begin{tabularx}{0.99\textwidth}{lcccccc}
    \toprule
    &\multicolumn{6}{c}{\textbf{Asset-Weighted Correlation}}  \\
    \midrule
    & \textbf{coef} & \textbf{std err} & \textbf{t} & \textbf{P$> |$t$|$} & \textbf{[0.025} & \textbf{0.975]}  \\
    \midrule
    \textbf{const} &       0.2004  &        0.025     &     7.991  &         0.000        &        0.150    &        0.250     \\
    \textbf{Securities Ratio}    &       0.5298  &        0.121     &     4.379  &         0.000        &        0.289    &        0.771     \\
    \bottomrule
    \end{tabularx}

    \begin{tabular}{lclc}
    \textbf{R-squared:}       &  0.204 & \textbf{  Durbin-Watson:     } &    1.133  \\ \\
    \end{tabular}
    
    \begin{tabularx}{0.99\textwidth}{lcccccc}
    \toprule
    &\multicolumn{6}{c}{\textbf{Asset-Weighted Correlation}}  \\
    \midrule
    & \textbf{coef} & \textbf{std err} & \textbf{t} & \textbf{P$> |$t$|$} & \textbf{[0.025} & \textbf{0.975]}  \\
    \midrule
    \textbf{const} &       0.2842  &        0.007     &    43.280  &         0.000        &        0.271    &        0.297     \\
    \textbf{HTM Ratio       }    &       0.1424  &        0.028     &     5.135  &         0.000        &        0.087    &        0.198     \\
    \bottomrule
    \end{tabularx}
    
    \begin{tabularx}{0.99\textwidth}{lclc}
    \textbf{R-squared:}       &  0.260 & \textbf{  Durbin-Watson:     } &    1.196  \\ \\ \\
    \end{tabularx}

    \begin{tabularx}{0.99\textwidth}{l|rrr}
    \toprule
    \textbf{Correlation Matrix} &  Securities Ratio &  HTM Ratio &  Asset Return\\
    &   &   &  Correlation\\    \midrule
    Securities Ratio    &   1.000000 &   0.861843 &      0.451278 \\
    HTM Ratio    &   0.861843 &   1.000000 &      0.509998 \\
    Asset Return Correlation&   0.451278 &   0.509998 &      1.000000 \\
    
    \bottomrule
    \end{tabularx}

\caption{Linear Dependency Between Asset Returns Correlation and Asset-Weighted Ratios}
\label{regressionResults}
\end{table}

The empirical evidence shows that U.S. banks have significantly increased their Held-To-Maturity (HTM) securities holdings post-COVID-19, particularly U.S. government securities, leading to potential hidden risks due to devaluation from interest rate hikes. The number of banks has declined, but asset consolidation among the largest banks has grown, with top banks holding a substantial portion of total assets. Leverage ratios remain stable, but the deposit outflow and high asset correlations among banks suggest elevated systemic risk. Overall, the increased HTM holdings and correlated asset returns could lead to systemic vulnerabilities, particularly during periods of financial stress. In the following section, we introduce a simple model to investigate the channel of systemic risk buildup and a way we could adopt to increase the system's resilience. 

\subsection{HTM Securities Graph and Statistics}

\begin{table}[ht]
    \centering
    \resizebox{1.0 \columnwidth}{!}{%
    \begin{tabular}{llrr}
    \hline
    {} &                                               \bb{Top Banks at March 2022} &  Total Assets(\$Billion) &  HTM Securities Ratio \\
    \hline
    1 &                                FIRST REPUBLIC BANK &       187.121 &                          0.886 \\
    2 &                                SILICON VALLEY BANK &       217.804 &                          0.792 \\
    3 &              BANK OF AMERICA, NATIONAL ASSOCIATION &      2513.619 &                          0.723 \\
    4 &             WELLS FARGO BANK, NATIONAL ASSOCIATION &      1764.272 &                          0.638 \\
    5 &  MORGAN STANLEY PRIVATE BANK, NATIONAL ASSOCIATION &       203.508 &                          0.603 \\
    \hline
    \\
    \end{tabular}
    }%
    
    \centering
    \resizebox{1.0 \columnwidth}{!}{%
    \begin{tabular}{llrr}
    \hline
    {} &                                  \bb{Top Banks at March 2023} &  Total Assets(\$Billion) &  HTM Securities Ratio \\
    \hline
    1 &                           UBS BANK USA &       113.867 &                          0.999 \\
    2 &                    FIRST REPUBLIC BANK &       232.944 &                          0.910 \\
    3 &  BANK OF AMERICA, NATIONAL ASSOCIATION &      2518.290 &                          0.818 \\
    4 &                   VALLEY NATIONAL BANK &        64.314 &                          0.749 \\
    5 &  MUFG UNION BANK, NATIONAL ASSOCIATION &        99.964 &                          0.737 \\
    \hline
    \end{tabular}
    }%

\caption{HTM Securities Ratio Rank}
\label{HTMratio Rank}
\end{table}

\begin{figure}[ht]
     \centering
     \begin{subfigure}[b]{0.6\textwidth}
         \centering
         \includegraphics[width=\textwidth]{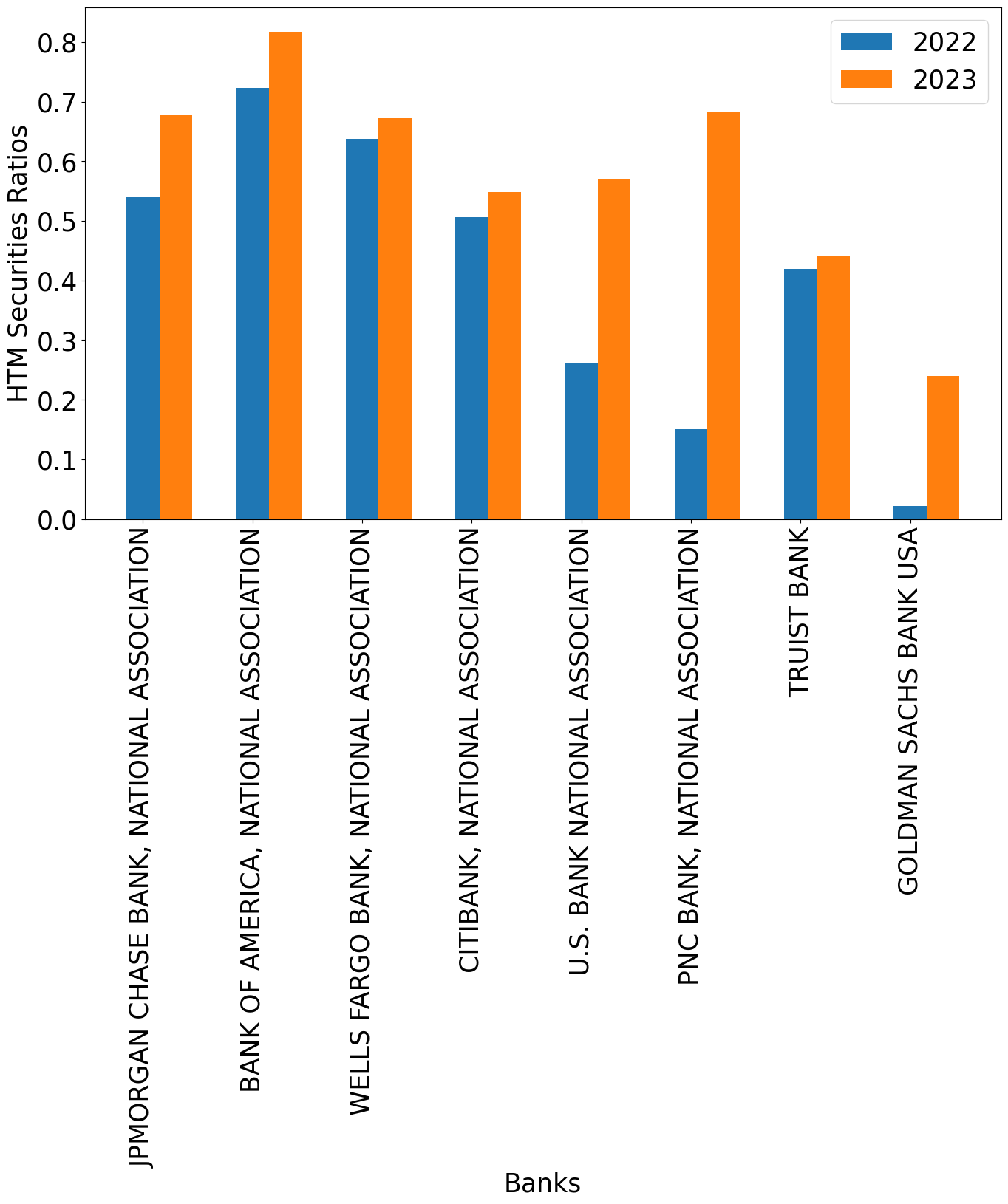}
         \caption{HTM Holdings of the 9 Largest Banks}
         \label{fig: HTMcompare}
     \end{subfigure}
     \hfill
     \begin{subfigure}[b]{0.6\textwidth}
         \centering
         \includegraphics[width=\textwidth]{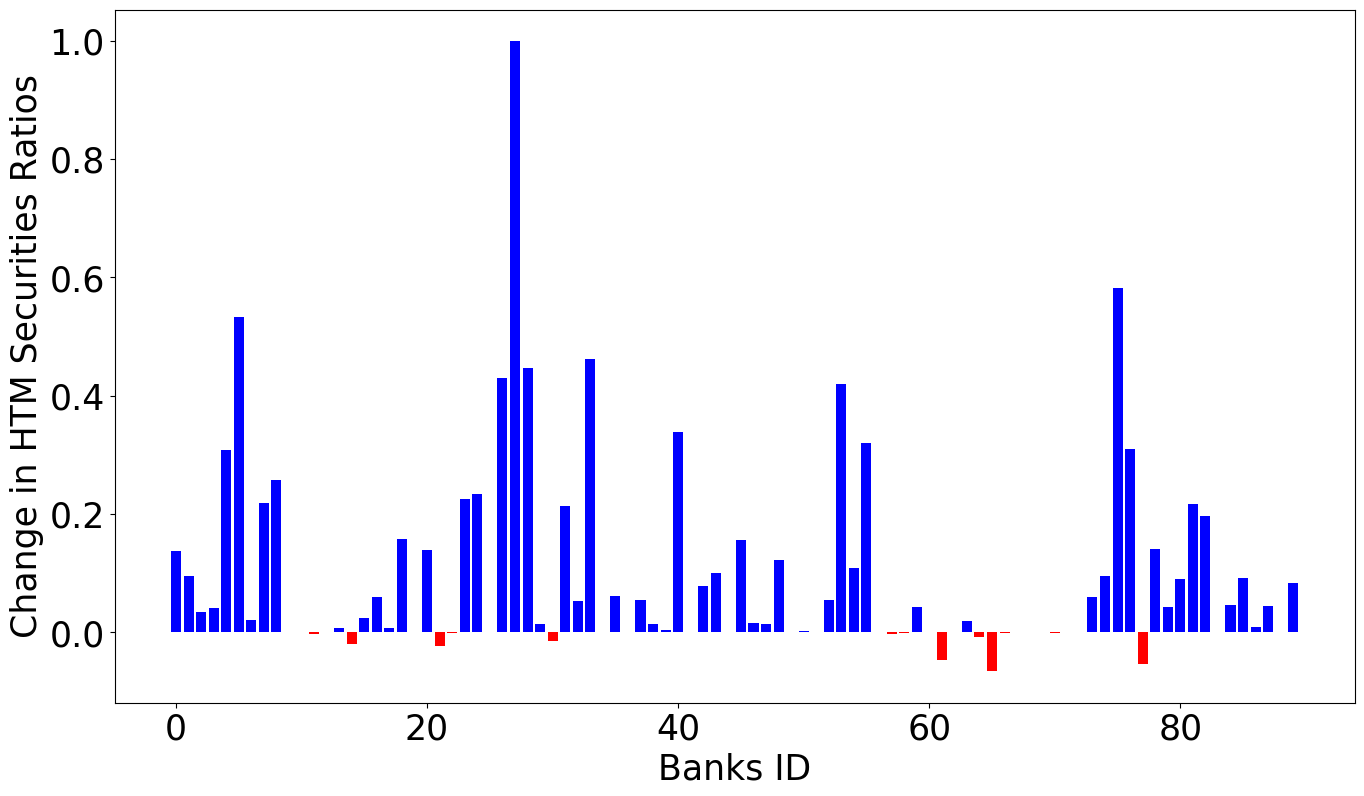}
         \caption{HTM Holdings Ratio increases from March 2022 to March 2023, Largest 100 banks}
         \label{fig: HTMchanges}
     \end{subfigure}
        \caption{Changes in HTM Securities Holdings}
        \label{fig:changesInHTMholding}
\end{figure}

\end{document}